\documentclass[aps,prb,notitlepage,twocolumn,10pt,noeprint]{revtex4-2}
\usepackage{dcolumn}
\usepackage{graphicx}
\usepackage[T1]{fontenc}
\usepackage[utf8]{inputenc}
\usepackage{multirow}
\usepackage[bookmarks=false,breaklinks,colorlinks]{hyperref}
\usepackage{mathtools}
\usepackage{siunitx}

\usepackage{lipsum}

\newcommand{\ie}{\textit{i}.\textit{e}.\ }

\newcommand{\dlambda}[1]{\frac{\partial #1(\lambda)}{\partial \lambda}}

\newcommand{\U}{U}

\newcommand{\hi}[1]{_{1{#1}}}
\newcommand{\lo}[1]{'_{1{#1}}}
\newcommand{\harm}{_0}

\newcommand{\uhi}{\U\hi{}}
\newcommand{\ulo}{\U\lo{}}
\newcommand{\uharm}{\U{\harm}}

\newcommand{\udhfull}{\left( \uhi - \uharm \right)}

\newcommand{\ebeta}[1]{e^{-\beta #1}}

\newcommand{\ave}[2]{\left< #1 \right>_{#2}}

\newcommand{\duhilo}{(\uhi-\ulo)}

\newcommand{\lambdalo}{{\lambda'}}

\newcommand{\var}{\operatorname{var}}

\newcommand{\varuhiharm}{\var(\uhi-\uharm)}
\newcommand{\gaussschwartz}{|\operatorname{cov}(X,Y)| \leq \sqrt{\operatorname{var}(X) \operatorname{var}(Y)}}
\newcommand{\weight}{\ebeta{\lambda \left(\uhi-\ulo\right)}}

\newcommand{\muhat}{\hat{\mu}}
\newcommand{\muhatprime}{\hat{\mu}'}
\newcommand{\muhatzero}{\hat{\mu}'_0}
\newcommand{\muhatone}{\hat{\mu}'_1}
\newcommand{\muhattwo}{\hat{\mu}'_2}
\newcommand{\muhatthree}{\hat{\mu}'_3}
\newcommand{\muhatfour}{\hat{\mu}'_4}
\newcommand{\cubegrid}[1]{$\mathrm{#1 \times #1 \times #1}$}

\newcommand{\dhatf}{\Delta \hat{F}}

\newcommand{\deltahatffep}{\dhatf_{11'}}
\newcommand{\deltahatffepn}{\dhatf_{11',\mathcal{N}}}
\newcommand{\FEPN}{{FEP-$\mathcal{N}$}}

\newcommand{\dftwo}{\dhatf(\muhattwo)}
\newcommand{\dffour}{\dhatf(\muhatfour)}

\usepackage{soul}

\begin{document}
\title{A statistical comparison of different approximate Hamiltonian-based 
anharmonic free energy estimators}

\author{E. Metsanurk} 
\email{erki.metsanurk@eesti.ee}
\affiliation{Department of Physics and Astronomy, Uppsala University, Box 516, S-75120 Uppsala, Sweden} 

\date{\today}

\begin{abstract}
Ensuring a satisfactory statistical convergence of anharmonic 
thermodynamic properties requires sampling of many atomic 
configurations, however the methods to obtain those necessarily produce 
correlated samples, thereby reducing the effective sample size and 
increasing the uncertainty compared to purely random sampling. In 
previous works procedures have been implemented to accelerate the 
computations by first performing simulations using an approximate
Hamiltonian which is computationally more efficient than the accurate 
one and then using various methods to correct for the resulting error. Those rely 
on recalculating the accurate energies of a random subset of 
configurations obtained using the approximate Hamiltonian thereby 
maximizing the effective sample size. This procedure can be particularly 
suitable for calculating thermodynamic properties using 
density-functional theory in which case the accurate and approximate 
Hamiltonians may be represented by parametrically suitably converged 
and non-converged ones. Whereas it is qualitatively known that there 
needs to be a sufficient overlap between the phase spaces of the 
approximate and the accurate Hamiltonians, the quantitative limits of 
applicability and the relative efficiencies of such methods is not well 
known. In this paper a statistical analysis is performed first 
theoretically and then quantitatively by numerical analysis. The 
sampling distributions of different free energy estimators are obtained 
and the dependence of their bias and variance with respect to 
convergence parameters, simulation times and reference potentials is 
estimated.
\end{abstract}

\maketitle

%%%%%%%%%%%%%%%%%%%%%%%%%%%%%%%%%%%%%%%%%%%%%%%%%%%%%%%%%%%%%%%%%%%%%%%%
\section{Introduction}
%%%%%%%%%%%%%%%%%%%%%%%%%%%%%%%%%%%%%%%%%%%%%%%%%%%%%%%%%%%%%%%%%%%%%%%%

Theoretical prediction of anharmonic thermodynamic properties of a 
material from first principles requires a fast and accurate method to 
sample the energies of relevant microstates of the atomic system. For 
instance, in order to obtain good estimates of the phase transition 
temperatures, the Gibbs free energies typically need to be determined 
to less than \SI{1}{\milli\electronvolt/atom} at high temperatures 
\cite{Bendick1982,Dinsdale1991,Grabowski2011a} which is a fraction of a 
percent of the total free energy. It has been proposed that density 
functional theory (DFT) \cite{Burke2012,Jones2015} based calculations 
are able to provide such a level of accuracy while taking into account 
the various contributions to the free energy, such as those from 
vibrational, electronic and magnetic excitations 
\cite{Palumbo2014,Moustafa2017}.

An integral part of any DFT calculation is verifying the convergence of 
the results with respect to various approximations, such as the number 
of explicitly treated electrons, truncation of the basis set of the 
wave functions, the number of k-points used to sample the Brillouin 
zone, smearing of the electronic states, stopping criteria for the 
optimization of the density and so on \cite{Wagner1998,Kratzer2019}. 
Apart from the lattice dynamics method, which provides an analytic 
expression \cite{Dove2011,Togo2015}, calculating the free energy of a 
crystal is done numerically by sampling atomic configurations using 
Monte Carlo (MC) or molecular dynamics (MD) simulations 
\cite{Frenkel,Rickman2002,Chipot2007} and subsequently transforming 
thermodynamic averages of energy differences to free energy 
differences. In order to keep the statistical error of the averages low 
at high temperatures, many simulation steps are needed, which is 
impractical to accomplish with highly converged DFT. Therefore it has 
become a standard practice to carry out the simulations using 
non-converged energies and forces, after which various techniques can 
be used to adjust the results to correct for the resulting error 
\cite{Vocadlo2002,Grabowski2009,Grabowski2011a, 
Duff2015,Moustafa2017,Sun2018,Rang2019}.

This systematic error can be thought of as consisting from the direct 
error in the computed energies due to the non-convergence and the 
indirect error due to the sampled structures being different from those 
that would be obtained from simulations using converged forces and 
energies. The former can be corrected simply by taking a smaller subset 
of the sampled configurations and recalculating the energies using 
parameters that ensure sufficient convergence. This is the idea behind 
up-sampled thermodynamic integration using Langevin dynamics (UP-TILD) 
\cite{Grabowski2009,Grabowski2011a} and its two-stage variation 
(TU-TILD) \cite{Duff2015}. In both cases the correction is applied to 
every step on the thermodynamic integration (TI) 
\cite{Frenkel1984,Frenkel,Kirkwood1935} path between a harmonic 
reference and DFT in the case of the former and an empirical potential 
and DFT when using the latter method.

This approximation is exact in the limit where the energy differences 
between converged and non-converged calculations do not depend on the 
atomic configuration, \ie remain constant throughout the simulation, 
which implies that the forces are not affected. In general, however, 
changing DFT parameters can have a considerable effect on the forces. 
In that case free energy perturbation (FEP) can be used to obtain an 
estimate of the error either by applying it directly \cite{Sun2018}, 
through its truncated cumulant expansion \cite{Vocadlo2002,Rang2019} or 
by reweighting ensemble averages \cite{Moustafa2017}. The latter 
approach, while strictly speaking not FEP, entails calculating the same 
exponential averages which can be significantly biased in the limit of 
a small dataset and can exhibit poor convergence with respect to the 
size of the dataset \cite{Wood1991, Zuckerman2002, Gore2003, 
Zuckerman2004, Wu2004, Wu2005, Jarzynski2006, Pohorille2010}. The 
problem is further exacerbated when FEP is performed unidirectionally 
\cite{Boresch2017} as is mostly the case when applying it to adjust the 
non-converged DFT results.

In this paper first a statistical analysis is performed theoretically, 
which is helpful to understand some aspects of the different free 
energy estimators, but is also limited since in general the 
distributions of energy differences are not known and have to be 
sampled numerically. In order to accelerate the latter, Spectral 
Neighbor Analysis Potentials (SNAP) \cite{Wood2018} are fitted to 
DFT-MD data of different levels of convergence. This allows for fast 
calculations of the sampling distributions of the energy differences 
between both the reference and the approximate Hamiltonians and between 
the accurate and the approximate Hamiltonians throughout the whole TI 
path. From these data the bias and the variance of the different
free energy estimators can be computed for different combinations
of DFT convergence parameters, expected simulation times and reference
potentials.

%%%%%%%%%%%%%%%%%%%%%%%%%%%%%%%%%%%%%%%%%%%%%%%%%%%%%%%%%%%%%%%%%%%%%%%%
\section{Theory and setup}
%%%%%%%%%%%%%%%%%%%%%%%%%%%%%%%%%%%%%%%%%%%%%%%%%%%%%%%%%%%%%%%%%%%%%%%%

%%%%%%%%%%%%%%%%%%%%%%%%%%%%%%%%%%%%%%%%%%%%%%%%%%%%%%%%%%%%%%%%%%%%%%%%
\subsection{Free energy estimators}\label{sec:theory}
%%%%%%%%%%%%%%%%%%%%%%%%%%%%%%%%%%%%%%%%%%%%%%%%%%%%%%%%%%%%%%%%%%%%%%%%
When the potential energy of a system depends on a parameter $\lambda$,
the partial derivative of the Helmholtz free energy with respect to
$\lambda$ is given by \cite{Frenkel}:

\begin{equation}\label{eq:dfdl}
    \left(\dlambda{F} \right)_{\lambda,NVT} = 
        \left< \dlambda{U} \right>_{\lambda,NVT}
\end{equation}

This can be used to calculate the free energy difference between two
states with different potential energies $\uhi$ and $\uharm$ by
choosing $\U(\lambda)$ as

\begin{equation}\label{eq:intpath}
\U(\lambda) = \lambda \uhi + (1-\lambda) \uharm
\end{equation}

and integrating both sides of Equation~\ref{eq:dfdl} from $\lambda=0$
to $\lambda=1$, giving

\begin{equation}\label{eq:lambdati}
    \left(F_1 - F_0\right)_{NVT} = 
        \int_0^1 \left<\uhi - \uharm\right>_{\lambda,NVT} d\lambda
\end{equation}

The free energy $F_1$ of any system with potential energy $\uhi$ can 
therefore by estimated by choosing a suitable reference potential 
$\uharm$ for which the free energy $F_0$ is known and integrating the 
potential energy difference between the systems on the path given by 
Equation~\ref{eq:intpath}. Common reference potentials for solids 
include uncoupled, \ie Einstein crystal, and coupled harmonic oscillators 
as in both cases the free energy can be calculated analytically.

In practice the integral can be evaluated by performing several 
equilibrium MC or MD simulations at different values of $\lambda$ and 
using, for example, a Gauss-Legendre quadrature or by fitting a function 
whose integral can be found analytically through the calculated points. 
Commonly a polynomial of a suitable degree is chosen while the 
$\lambda$ values can be either equidistant or not 
\cite{Shyu2009,Jorge2010}. In some cases more sophisticated 
trigonometric functions have been used in order to get a better fit 
compared to a polynomial with the same number of parameters 
\cite{Grabowski2011a}.

Regardless of the chosen integration method, when DFT-MD is used to 
estimate the free energy difference, both the errors in the potential 
energy $\uhi$ and forces ${\vec{f\hi{}}=-\nabla \uhi}$ due to chosen 
approximations (DFT convergence) and the uncertainty of the ensemble 
average at each $\lambda$ (statistical convergence) have to be kept 
small enough to ensure adequate accuracy of the results. 

The natural way to estimate $\mu(\lambda) = \ave{\uhi-\uharm}{\lambda}$,
is to take the arithmetic mean of the samples obtained from MD using
potential $\U(\lambda)$
\begin{equation}\label{eq:muhat}
\muhat(\lambda) = \frac{1}{N} \displaystyle\sum_{i=1}^{N} \left(\uhi{_i}-\uharm{_i}\right)
\end{equation}
however, due to autocorrelation of the samples, this can be quite inefficient
compared to random sampling, since the variance of $\muhat$ does
not decrease in proportion with the sample size $N$, but the effective
sample size, $N_\mathrm{eff}$ which can be significantly smaller.

If $\uhi$ in Equation~\ref{eq:intpath} is replaced with
another potential $\ulo$, such that

\begin{equation}\label{eq:approximateU}
\U'(\lambda) = \lambda \ulo + (1-\lambda) \uharm
\end{equation}

then if

\begin{subequations}\label{eq:mulambdap}
 \begingroup
\begin{align}
\mu'(\lambda) &= \frac{ 
        \ave{\udhfull \weight}{\lambdalo}
    }{\ave{\weight}{\lambdalo}} \label{eq:reweigh_a}\\
    &= 
    \ave{\uhi-\uharm}{\lambdalo} 
+ \frac{\operatorname{cov}\left(\uhi-\uharm, \ebeta{\lambda (\uhi-\ulo)} \right)
_{\lambdalo}}{\ave{\ebeta{\lambda \left(\uhi-\ulo\right)}}{\lambdalo}} \label{eq:reweigh_b}
\end{align}
\endgroup
\end{subequations}

it follows that

\begin{equation}
\mu'(\lambda) = \mu(\lambda)
\end{equation}

The latter can be easily shown, since for any property $A$
that depends on the coordinates and momenta of the atoms,

\begin{equation}
\left< A \right>_{\lambda} = \frac{ \ave{A \ebeta{[\U(\lambda) - \U'(\lambda)]}}{\lambdalo} }{ \ave{\ebeta{[\U(\lambda) - \U'(\lambda)]}}{\lambdalo} }
\end{equation}

where the subscripts $\lambda$ and $\lambdalo$ denote that the
potential energy of the ensemble is $\uhi$ and $\ulo$ respectively.

In order to analyze the advantage of estimating $\mu'$ over 
$\mu$, we will first consider the case when 

\begin{equation}\label{eq:zero_cov}
  \operatorname{cov} \left(\uhi-\uharm, \ebeta{\lambda (\uhi-\ulo)} 
  \right)_\lambdalo = 0 
\end{equation} 
%\begin{equation}\label{eq:zero_cov}
%\frac{\operatorname{cov}\left(\uhi-\uharm, \ebeta{\lambda (\uhi-\ulo)} \right)
%_{\lambdalo}}{\ave{\ebeta{\lambda \left(\uhi-\ulo\right)}}{\lambdalo}} = 0
%\end{equation} 

which, although not explicitly shown 
in the original work, is the approximation behind the UP-TILD method 
\cite{Grabowski2009}.
The corresponding unbiased estimator of $\mu'$ is then

\begin{equation}
\begin{aligned}
\muhatzero(\lambda) &= \frac{1}{N} \displaystyle\sum_{i=1}^{N} \left(\uhi{_i}-\uharm{_i}\right) \\
&= \frac{1}{N} \displaystyle\sum_{i=1}^{N} \left(\U\lo{i}-\uharm{_i}\right) + \frac{1}{N} \displaystyle\sum_{i=1}^{N} \left(\uhi{_i}-\U\lo{i}\right)
\end{aligned}
\end{equation}

Note that whereas the subscripts $\lambdalo$ are omitted from this and
the following estimators, it is implied that the samples are obtained
using the potential $\ulo$.

Using $\muhatzero$ does not provide any possible improvement in 
efficiency over using $\muhat$, since it entails first performing the 
simulation using potential $\U'(\lambda)$ followed by recalculating the 
energy of every sample using $\uhi$. However, as pointed out above, for a 
given sample variance it is equivalent to either take the average $N$ 
correlated samples or $N_\mathrm{eff}$ random samples. The simplest way 
to do the latter is to calculate the mean of every $k$-th sample such 
that the autocorrelation function for lag $k$ has decreased to a 
sufficiently small value. This results in estimators

\begin{equation}\label{eq:muhatuptild_onlyrecalc}
\begin{aligned}
\muhatone(\lambda) &= \frac{k}{N} \displaystyle\sum_{i=1}^{N/k} \left(\uhi{_{ki}}-\uharm{_{ki}}\right)
\end{aligned}
\end{equation}

and

\begin{equation}\label{eq:muhatuptild}
\begin{aligned}
\muhattwo(\lambda) &= \frac{1}{N} \displaystyle\sum_{i=1}^{N} \left(\U\lo{i}-\uharm{_i}\right) +  \frac{k}{N} \displaystyle\sum_{i=1}^{N/k} \left(\uhi{_{ki}}-\U\lo{{ki}}\right)
\end{aligned}
\end{equation}

which can be faster to evaluate than $\hat{\mu}$, assuming that it is 
faster to calculate $\ulo$ compared to $\uhi$. The difference between 
Equations~\ref{eq:muhatuptild_onlyrecalc} and \ref{eq:muhatuptild} is 
whether all of the original energies are taken into account or only the 
ones corresponding to the recalculated configurations. This 
has no direct effect on the computational time, but the variances of 
the estimators can differ. This will be investigated in more detail
in Section~\ref{sec:muhat_comparison}.

In order for $\muhatone$ and $\muhattwo$ to be unbiased, a sufficient 
condition for $\ulo$ is

\begin{equation}\label{eq:zero_var}
\operatorname{var}\duhilo_\lambdalo = 0 \quad \forall \lambda \in [0,1]
\end{equation}

That is a stronger requirement than that of Equation~\ref{eq:zero_cov}, 
and when true, means that $k=N$ can be taken in the second term of 
$\muhattwo$, \ie only a single recalculation is 
needed. In practice, the variance does not need to be exactly zero. If 
Equation~\ref{eq:zero_cov} is not satisfied, then 
$\muhattwo$ is a biased and inconsistent estimator of 
$\mu'$, but if the bias is smaller than the required 
accuracy, the approximation can still be used. Moreover, at $\lambda = 
0$, Equation~\ref{eq:zero_cov} is always true, regardless of how large 
$\operatorname{var}\duhilo_{0'}$ is, since the potential energy $\U'(0)$ 
(in Equation~\ref{eq:approximateU})
does not depend on $\uhi$. Taking all of the above into account
it follows that $\ulo$ should be a close 
approximation of $\uhi$ up to a constant and a typical choice for that 
is non-converged DFT.

Without any approximations, $\mu'$ can be estimated by using either

\begin{equation}
\muhatthree(\lambda) = \frac{\displaystyle\sum_{i=1}^{N/k} \left(\uhi{_{ki}}-\uharm{_{ki}}\right) 
\ebeta{\lambda \left(\uhi{_{ki}}-\U\lo{ki}\right)}}{\displaystyle\sum_{i=1}^{N/k} \ebeta{\lambda \left(\uhi{_{ki}}-\U\lo{ki}\right)}}
\end{equation}

or

\begin{equation}
\begin{aligned}
\muhatfour(\lambda) &= \frac{1}{N} \displaystyle\sum_{i=1}^{N} \left(\U\lo{i}-\uharm{_i}\right)
  - \frac{k}{N} \displaystyle\sum_{i=1}^{N/k} \left(\U\lo{ki}-\uharm{_{ki}}\right) \\
& + \frac{\displaystyle\sum_{i=1}^{N/k} \left(\uhi{_{ki}}-\uharm{_{ki}}\right) 
\ebeta{\lambda \left(\uhi{_{ki}}-\U\lo{ki}\right)}}{\displaystyle\sum_{i=1}^{N/k} \ebeta{\lambda \left(\uhi{_{ki}}-\U\lo{ki}\right)}} \\
& = \muhattwo+\frac{\displaystyle\sum_{i=1}^{N/k} \left(\uhi{_{ki}}-\uharm{_{ki}}\right) 
\ebeta{\lambda \left(\uhi{_{ki}}-\U\lo{ki}\right)}}{\displaystyle\sum_{i=1}^{N/k} \ebeta{\lambda \left(\uhi{_{ki}}-\U\lo{ki}\right)}} \\
 &- \frac{k}{N} \displaystyle\sum_{i=1}^{N/k} \left(\uhi{_{ki}}-\uharm{_{ki}}\right)
\end{aligned}
\end{equation}

the difference being that at the limit of Equation~\ref{eq:zero_var} 
the former approaches $\muhatone$ and the latter $\muhattwo$ thereby 
making use of all of the available data.

Both $\muhatthree$ and $\muhatfour$ are biased, but consistent 
estimators of $\mu'$ meaning that as the number of samples $N$ 
goes to infinity, the bias approaches $0$. However, for small $N$ and 
large $\operatorname{var}\duhilo_\lambdalo$ it is possible that the 
uncertainty due to the bias and the variance of the exponential 
ensemble averages \cite{Jarzynski2006,Boresch2017,Ryde2017} is so large 
that estimating $\mu'$ instead of $\mu$ might not 
provide any improvement or even be less efficient. An exception to that 
is when ${(\uhi-\uharm)_\lambdalo}$ and ${(\uhi-\ulo)_\lambdalo}$ follow a bivariate normal 
distribution. In that case Equation~\ref{eq:mulambdap} simplifies to

\begin{equation}
\mu'_{\mathcal{N}}(\lambda) = \ave{\uhi-\uharm}{\lambdalo} -\lambda \beta \operatorname{cov} \left(\uhi-\uharm, \uhi-\ulo\right)_\lambdalo
\end{equation}

which can be estimated without bias since the sample covariance
is an unbiased estimator of the ensemble covariance.

For any estimator $\muhat'$ disrobed above the free energy difference
$F_1-F_0$ is estimated as

\begin{equation}
\dhatf (N,k) = \int_0^1 \muhat' (\lambda; N,k) d\lambda
\end{equation}

It is noteworthy that even if $\muhat'$ is biased at almost every
$\lambda$, it is possible that $\dhatf$ is unbiased, if
the bias integrates to $0$, for example when using $\muhattwo$ and

\begin{equation}\label{eq:zero_int}
\int_0^1 \operatorname{cov} \left(\uhi-\uharm, \ebeta{\lambda (\uhi-\ulo)} \right)_\lambdalo d\lambda = 0
\end{equation}

\begin{figure*}
    \includegraphics[width=\textwidth]{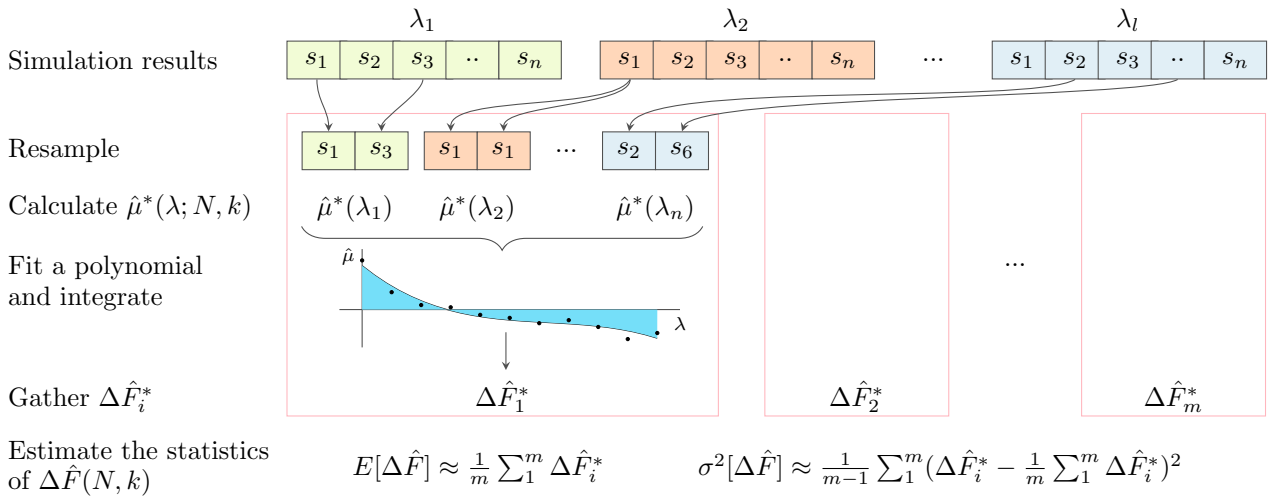}
    \caption{\label{fig:samplingdist} A scheme for obtaining the
    sampling distribution and the statistics of a free energy estimator. 
    Refer to the text for a detailed explanation.}
\end{figure*}

Based on the computational results in this work, it can be hypothesized 
that for a given $\uhi$ and $\ulo$ it is in principle possible to find 
$\uharm$ such that Equation~\ref{eq:zero_int} is true, however doing 
that might be impractical. Nevertheless the choice of $\uharm$ is 
important, since for random variables $X$ and $Y$ 
\begin{equation}\label{eq:gaussschwartz} \gaussschwartz \end{equation} 
\ie minimizing $\varuhiharm_\lambdalo$ will reduce the bias, especially 
when $\lambda$ is close to $1$. This argument also applies to other 
estimators of $\mu'$. In this paper the significance of that is studied 
by comparing several different reference potentials.

The integral of the first term of both $\muhattwo$ and $\muhatfour$ estimates

\begin{equation}\label{eq:df_uncorrected}
\int_0^1 \left<\ulo-\uharm\right>_\lambdalo d\lambda = F_1' -F_0
\end{equation}

Therefore the integral of the other terms estimates
\begin{equation}
F_1-F_0-(F_1'-F_0) = F_1-F_1'
\end{equation}

\ie the free energy difference between states with potentials $\uhi$ 
and $\ulo$. Varying $\lambda$ from 0 to 1, however switches the 
potential from $\uharm$ to $\ulo$. It can be argued that except when 
Equation~\ref{eq:zero_int} is true, as described above, there is no 
obvious reason to expect that estimating the free energy difference 
$F_1-F_1'$ via a path that does not directly connect the corresponding 
states is more efficient than that which does. Typically $\uharm$ is a 
significantly worse approximation of $\uhi$ than $\ulo$, in which case 
the samples of $\uhi-\ulo$ near $\lambda=0$ provide much less 
information about $F_1-F_1'$ than those near $\lambda=1$ so it might be 
advantageous to just gather more samples at that endpoint of the path.

Moreover, since successful application of 
either $\muhattwo$ or $\muhatfour$ requires 
$\operatorname{var}\duhilo_\lambdalo$ to be as small as possible
anyway, it could be more reasonable to calculate $F_1-F_1'$ directly using FEP from

\begin{equation}
 \label{eq:fep}
 \begin{aligned}
\Delta F_{11'} = F_1-F'_1 = - \beta^{-1} \ln \left< e^{-\beta (U_1-U'_1)}\right>_{\lambda'=1}
\end{aligned}
\end{equation}

When $(\uhi-\ulo)_{\lambdalo=1}$ has a Gaussian distribution, a second order cumulant 
expansion of FEP (\FEPN)can be used as

\begin{equation}
\label{eq:fepvar}
  \Delta F_{11'} = F_1-F'_1 = \left< \uhi - \ulo\right>_{\lambda'=1}-
        \frac{\beta}{2} \operatorname{var}(\uhi-\ulo)_{\lambda'=1}
\end{equation}

which has the advantage of converging much faster than the full FEP 
\cite{Ryde2017} and being able to estimate it without bias.
When estimating $F_1-F'_1$, then similarly to $\muhat'$ it is advantageous
to take the averages over uncorrelated samples. If every
k-th sample is used, the corresponding estimators are

\begin{equation} \label{eq:hatfep}
\deltahatffep = - \beta^{-1} \ln \frac{k}{N} \sum_{i=1}^{N/k} e^{-\beta \Delta U_{ki}}
\end{equation}

and 

\begin{equation}\label{eq:hatfepvar}
\begin{aligned}
\deltahatffepn &= \frac{k}{N} \displaystyle\sum_{i=1}^{N/k} \Delta U_{ki}  \\
&-        \frac{\beta}{2} \frac{1}{N/k-1} \displaystyle \sum_{i=1}^{N/k} \left( \Delta U_{ki} - \overline{\Delta U_k} \right)^2
\end{aligned}
\end{equation}

where $\Delta U_{ki}  = \uhi{_{ki}}-\U\lo{{ki}}$ and
$\overline{\Delta U_k} = (k/N) \sum_{i=1}^{N/k} \Delta U_{ki} $,
both sampled at $\lambda=1$ with $\ulo$.

%%%%%%%%%%%%%%%%%%%%%%%%%%%%%%%%%%%%%%%%%%%%%%%%%%%%%%%%%%%%%%%%%%%%%%%%
\subsection{Uncertainties of the estimators}
%%%%%%%%%%%%%%%%%%%%%%%%%%%%%%%%%%%%%%%%%%%%%%%%%%%%%%%%%%%%%%%%%%%%%%%%

For the given potentials $\uharm$, $\uhi$ and $\ulo$, \ie the
reference, accurate and the approximate one respectively, all the described 
estimators of $\Delta F$ are expected to differ in accuracy and 
precision. Whereas it might be possible to quantify those by theoretical 
means, starting from the $\lambda$-dependent multivariate distributions 
of the potential energies together with a model of time correlations, in 
practice it is easier to obtain the sampling distributions of the 
different $\dhatf$ and their dependence on $N$ and $k$ using 
simulations. This also avoids making any assumptions about the underlying 
distributions or whether other approximations such as the central limit 
theorem can be applied.

In order to get a sufficiently converged sampling distribution, enough 
values of $\dhatf$ have to be calculated. If the number of 
those is $m$, then that also requires $m$ values of $\muhatprime$ at 
$n_\lambda$ values of $\lambda$, which results in a total number of $m 
n_\lambda$ simulations. However, because the values of $\muhatprime$ 
at any $\lambda$ are uncorrelated with the values at any other 
$\lambda$, the aforementioned calculations can in fact provide 
$m^{n_\lambda}$ estimates of $\dhatf$. If there are more than 
$m$ values of $\muhatprime$ to pick $m$ out of, the number of estimates 
increases even more and even for modest values of $m$ and $n_\lambda$ 
it becomes vast. A small subset of those, in this work obtained by random 
sampling, can be used to estimate the sampling distribution of $\Delta 
\hat{F}$.

The simplified overview of the method used here is explained in 
Figure~\ref{fig:samplingdist}. First, at each $\lambda$ a series of MD 
simulations, differing only by the initial conditions, are run in order 
to precompute sufficiently representative samples of $\uharm$, $\uhi$ 
and $\ulo$. Alternatively a single long simulation could be performed, 
however it has been shown that the former helps with faster spanning of 
the phase space and improving parallelization \cite{Hellman2012}.

\begin{table*}
\caption{\label{tab:dft_parameters} DFT parameters}
\begin{ruledtabular}
\bgroup
\def\arraystretch{1.2}
\begin{tabular}{lccccc}
Parameter  & DFT-0 & DFT-1 & DFT-2 & DFT-3 & DFT-4 \\
\hline
Cutoff energy (\si{\electronvolt}) & 400.0 & 223.1 & 167.3 & 223.1& 167.3 \\
K-points ($\Gamma$-centered) &  \cubegrid{4} & \cubegrid{2} & \cubegrid{2} & \cubegrid{1} & \cubegrid{1} \\
FFT grid & \cubegrid{64} & \cubegrid{36} & \cubegrid{32} & \cubegrid{36} & \cubegrid{32} \\
Fine FFT grid & \cubegrid{128} & \cubegrid{72} & \cubegrid{48} & \cubegrid{72} & \cubegrid{48} \\ 
Projection space & reciprocal & real & real & real & real\\
Stopping tolerance & $1.0^{-6}$ & $1.0^{-6}$ & $1.0^{-6}$ & $1.0^{-6}$ & $1.0^{-4}$ \\
XC-functional & \multicolumn{5}{c}{Perdew-Burke-Ernzerhof generalized gradient approximation}  \\
Electrons per atom & \multicolumn{5}{c}{6} \\
Occupancy smearing & \multicolumn{5}{c}{Fermi-Dirac, $\sigma = \SI{0.318}{\electronvolt}$} \\
%\hline
Relative speed & 1 & 16 & 28 & 50 & 95
\end{tabular}
\egroup
\end{ruledtabular}
\end{table*}

In the next step $m$ estimates of $\Delta F$ are obtained. This is 
achieved by first choosing the number of timesteps $N$ and the constant 
$k$. Then the estimators $\muhat (\lambda; N,k)$ are applied to the 
data which are resampled with replacement from the set of precomputed 
short simulations such that the total number of timesteps is $N$. For 
example, if the number of timesteps in a single simulation was $5000$ 
and $N = 20000$, the results of 4 randomly chosen simulations are 
combined together. Since the simulations are independent, the 
correlations in the time series which reduce the effective sample size 
are trivially preserved. In order to simplify the analysis, $N$ is here 
taken to be constant with respect to $\lambda$. Next a polynomial is 
fitted through the obtained $\muhat(\lambda)$ data points. The degree 
of the polynomial is chosen such that the leave-one-out 
cross-validation score is minimized in order to avoid overfitting. The 
integral of the polynomial from $\lambda=0$ to $\lambda = 1$ is then 
stored.

In the third step the expected value and variance of $\dhatf$
can be obtained from the estimates of $\Delta F$ computed in the
previous step. For a good estimate of the bias, the reference $\Delta F$
is computed from the integral of Equation~\ref{eq:muhat} with all
the available data.

As mentioned, the scheme in Figure~\ref{fig:samplingdist} is a slight 
simplification as in practice the different means in $\muhat^*$ were 
calculated, fitted to a different order polynomials and integrated 
separately. This allows for analyzing the contributions of the 
different parts to the uncertainty of $\hat{F}$. In addition, since $k$ 
can be relatively large (at the limit equal to $N$), in order to better 
utilize the available data and converge the sampling distributions of 
$\hat{F}$ faster, each of the simulations was resampled with random 
starting offset between the first and the k-th timestep.

This approach of obtaining the sampling distribution closely resembles 
the bootstrap method \cite{Efron1979}. The main difference here is that 
the resampled sample sizes are considerably smaller than the original 
dataset which results in more accurate sampling distributions. In other 
words, when enough data have been precalculated, the error made by 
sampling from those instead of the canonical distribution becomes 
insignificant. This can be checked, for example, by dividing the 
precalculated data into multiple chunks, calculating the sampling 
distributions from each of those separately, and verifying that the 
results do not vary appreciably.

%%%%%%%%%%%%%%%%%%%%%%%%%%%%%%%%%%%%%%%%%%%%%%%%%%%%%%%%%%%%%%%%%%%%%%%%
\subsection{Potentials}
%%%%%%%%%%%%%%%%%%%%%%%%%%%%%%%%%%%%%%%%%%%%%%%%%%%%%%%%%%%%%%%%%%%%%%%%

Several useful conclusions about the statistics of the different 
$\hat{F}$ could likely be drawn by performing numerical simulations 
using any reasonable set of $\uharm$, $\ulo$ and $\uhi$ even without 
any DFT calculations. For example, a simple analytical potential could 
be taken as $\uhi$ and small perturbations made to its parametrization 
in order to obtain $\ulo$. Although being simple to implement and very 
fast, this would not provide much quantitative information about 
realistic problems which could solved using DFT. On the other hand, 
even with the resampling method described above, resources are limited 
to calculate everything using DFT-MD. A compromise can be made by 
performing the analysis with less computationally expensive potentials 
fitted to DFT. In this work, the choice is quadratic SNAP 
\cite{Wood2018} due to its good and tunable accuracy and ease of 
fitting, while being orders of magnitude faster than DFT.

A 54-atom supercell of BCC tungsten with lattice parameter of 
\SI{3.242}{\angstrom} was chosen as the system to be investigated. The 
training data were calculated using \textsc{vasp}
\cite{Kresse1993,Kresse1994a,Kresse1996,Kresse1996a} with 
projector-augmented wave method \cite{Kresse1999}. The convergence 
parameters used for the reference (\mbox{DFT-0}) and the successively 
worsely converged (\mbox{DFT-1} to \mbox{DFT-4}) electronic structure 
calculations are given in Table~\ref{tab:dft_parameters}. The canonical 
ensemble was sampled at \SI{3687}{\kelvin} using Langevin dynamics with 
\SI{10}{\pico\second^{-1}} friction coefficient and 
\SI{5}{\femto\second} timestep.

As shown in Table~\ref{tab:dft_parameters}, the speedup achieved by 
using non-converged instead of converged \mbox{DFT-MD} was between one 
and two orders of magnitude, with the fastest and slowest calculation 
taking 152 and 14457 core-seconds per timestep respectively. 
Recalculating the energies using \mbox{DFT-0} for structures sampled by 
non-converged \mbox{DFT-MD} was however two times slower, 28230 
core-seconds per timestep. This is due to the high correlation between 
subsequent samples in MD allowing for prediction of the wavefunctions 
which results in faster convergence of the electronic structure. Given 
that the recalculation is typically done for uncorrelated samples, this 
sort of prediction is not applicable.

% DFT-0  14457 cs, DFT-1 910 cs, 508 cs, 289 cs, 152 cs

The fitting was done using \textsc{fitsnap} \cite{fitsnap}. The training data 
consisted of energies and forces of 5374 and 7000 configurations for 
\mbox{DFT-0} and the non-converged DFT respectively. The maximum order of the bispectrum 
components was set to $J_\textrm{max} = 4$, cutoff distance to $R_\textrm{max} = 
\SI{4.8}{\angstrom}$ and the maximum latitude for remapping neighbor positions 
to $\theta_0^\textrm{max} = 0.99363 \pi$. In order to handle short atomic 
distances, Ziegler-Biersack-Littmark potential was added with smooth 
transition to zero between $R_\textrm{zbl,i} = \SI{4.0}{\angstrom}$ and 
$R_\textrm{zbl,o} = \SI{4.8}{\angstrom}$.

\begin{figure}
    \includegraphics[width=\columnwidth]{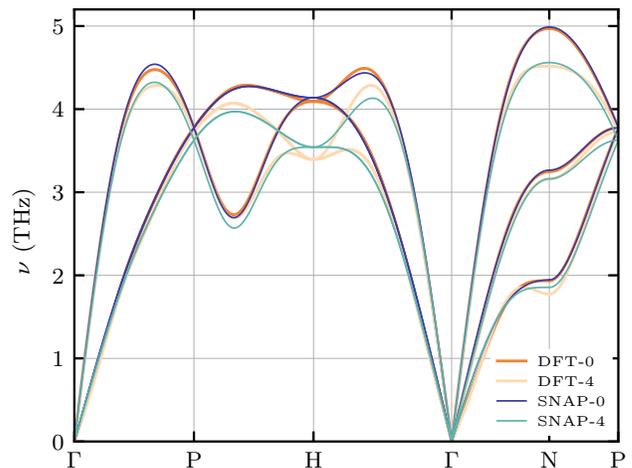}
    \caption{\label{fig:dft_snap_tdep_phonon} High temperature phonon
    dispersions of DFT and SNAP. The force constant matrices were fitted
    to forces and displacements obtained from \SI{3687}{\kelvin} molecular
    dynamics simulations. }
\end{figure}

Among all of the fitted potentials, the lowest root mean square errors 
of energy per atom and force components were 
\SI{2.6}{\milli\electronvolt} and \SI{0.13}{\electronvolt/\angstrom} 
for \mbox{SNAP-0} (fitted to \mbox{DFT-0}) and the largest errors were 
\SI{4.4}{\milli\electronvolt} and \SI{0.22}{\electronvolt/\angstrom} 
for \mbox{SNAP-4} (fitted to \mbox{DFT-4}). This is also illustrated on 
Figure~\ref{fig:dft_snap_tdep_phonon}, which depicts the phonon 
dispersions of the harmonic temperature-dependent effective potentials 
(TDEP) \cite{Hellman2011} fitted to the high temperature MD data for 
DFT and the corresponding SNAP. There is very little difference between 
\mbox{SNAP-0} and \mbox{DFT-0} and whereas the error is slightly larger 
between \mbox{SNAP-4} and \mbox{DFT-4}, the former is able to 
adequately reproduce the overall decrease in the effective frequencies 
caused by non-converged DFT. 

The average computational cost of SNAP 
during the molecular dynamics simulations was 0.02 core-seconds per 
timestep, \ie 4 to 6 orders of magnitude faster than DFT.

\begin{figure}
    \includegraphics[width=\columnwidth]{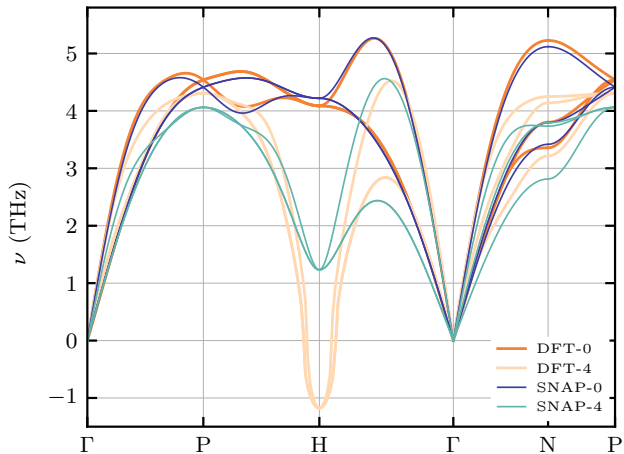}
    \caption{\label{fig:dft_snap_0k_phonon} Comparison of the \SI{0}{\kelvin} phonon
    dispersions of DFT and SNAP. }
\end{figure}

%\lipsum[2]

%%%%%%%%%%%%%%%%%%%%%%%%%%%%%%%%%%%%%%%%%%%%%%%%%%%%%%%%%%%%%%%%%%%%%%%%
\subsection{Reference potentials and switching calculations}
%%%%%%%%%%%%%%%%%%%%%%%%%%%%%%%%%%%%%%%%%%%%%%%%%%%%%%%%%%%%%%%%%%%%%%%%

Almost any potential can be used as $\uharm$, given its free energy is 
known or can be calculated by another simulation. A simple and easily 
obtainable choice is the harmonic approximation (HA) of the target 
potential whose free energy has to be estimated. However, because here 
$\ulo$ is used for sampling and $\uhi$ for recalculations, it is not 
clear whether the HA of one should be preferred over that of the other. 
One one hand, the closer the reference potential is to the target, the 
more efficient the $\lambda$-switching calculation, which suggests 
using the HA of $\ulo$. On the other hand, the configurations sampled 
with a potential more similar to that which is used for recalculating 
the energies, might make the computationally more expensive 
recalculations more efficient, suggesting the use of the HA of $\uhi$.

An exception to when the HA could be used as a reference is when some 
of the phonons have imaginary frequencies since then the free energy is 
not real-valued. This happens to be the case with DFT-4 near the 
H-point as shown on Figure~\ref{fig:dft_snap_0k_phonon}. Despite that 
the SNAP training data consisted of only high temperature MD forces and 
energies, the \SI{0}{\kelvin} phonon dispersions were still adequately 
reproduced while not containing any imaginary frequencies. Therefore 
the HAs of the fitted potentials were used instead of the DFT ones.

With increasing anharmonicity, the \SI{0}{\kelvin} HA is expected by 
definition to become a successively worse reference potential in terms 
of efficiency. An improved reference can be obtained by fitting another 
potential that can take the anharmonicity either implicitly or 
explicitly into account. In this work two types of such potentials were 
used. The first ones were the effective harmonic ones described above. 
The advantage of those is that the reference free energy is known 
analytically and no extra simulations are needed. The second ones were 
simpler linear SNAPs (from here on referred to as POT in order to avoid 
confusing it with the SNAPs approximating DFT) \cite{Thompson2014} with 
reduced maximum order of bispectrum components ($J_\textrm{max} = 3$) 
compared to the DFT-based potentials and the other parameters 
unmodified. Using those the reference free energy needs to be 
calculated separately and whereas this requires setting up simulations 
and performing additional analysis, the improvement in efficiency can 
be significant \cite{Duff2015}. As with the HA reference, since it is 
not immediately obvious whether the fitted potentials should be based 
on $\ulo$ or $\uhi$, both were compared in this work. Since using the 
latter to directly sample the configurations can be computationally 
expensive, it might instead be necessary to use the recalculated 
energies and forces for fitting, the effect of which is also 
investigated.

\begin{figure}
    \includegraphics[width=8cm]{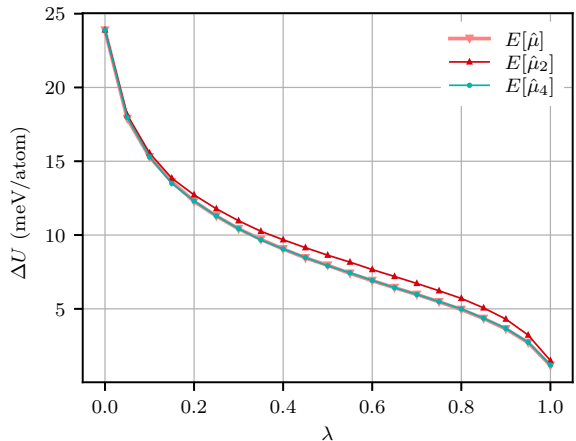} 
    \caption{\label{fig:switching} The $\lambda$-dependent expected values of $\muhat$,
    $\muhattwo$ and $\muhatfour$. HA was used as $\uharm$ and SNAP-4 as $\ulo$.
    The expected value was estimated by averaging over all of the available
    data with $k=1$.}
\end{figure}

All the simulations were performed using 
\textsc{lammps}~\cite{Plimpton1997}. In order to shorten the 
equibliration time, the initial positions and velocities were randomly 
sampled from distributions determined by the HA of the target potential 
\cite{Hellman2012}. The canonical ensemble was sampled at \SI{3687}{K} 
by a Generalized Langevin Equation thermostat 
\cite{Ceriotti2009,Ceriotti2010a} with a timestep of \SI{1}{fs}. The 
drift matrix of the thermostat was generated for optimal sampling in 
frequency range between 0.07 and \SI{7}{THz}. Each simulation consisted 
of \SI{12}{ps} out of which the first \SI{2}{ps} was equibliration. 
Every 10th configuration was stored and later recalculated using, 
resulting in 1000 $\uhi$, $\ulo$ and $\uharm$ values per simulation. 
The total number of simulations at each $\lambda$ (21 equidistant 
values between 0 and 1) was 200 for the HA and TDEP references, and 60 
for the POT ones. An example of the results is shown in 
Figure~\ref{fig:switching}.

%\cite{Martyna1992}

Due to the combination of the chosen geometry of the simulation box, 
lattice type and high temperature, occasionally the whole crystal 
rotated relative to the box to another symmetry-equivalent 
configuration. Whereas this did not pose a problem to the SNAP 
potentials, due to the fixed reference positions the displacements for 
evaluating the harmonic energies and forces became erroneous. In this 
case the results of the simulation were discarded and another one with 
different initial conditions was performed. Since the number of those
was relatively low due to the short simulation times, about 1\% at
$\lambda=1$ and none at $\lambda = 0$ since the harmonic potential
constrains that type of rotation, the effect on the results is expected
to be minimal. Another solution would be to use a supercell for which
such rotations are not possible, such as $4 \times 4 \times 4$ instead
of $3 \times 3 \times 3$.

%%%%%%%%%%%%%%%%%%%%%%%%%%%%%%%%%%%%%%%%%%%%%%%%%%%%%%%%%%%%%%%%%%%%%%%%
\section{Results}
%%%%%%%%%%%%%%%%%%%%%%%%%%%%%%%%%%%%%%%%%%%%%%%%%%%%%%%%%%%%%%%%%%%%%%%%

\subsection{The effect of reference systems}\label{sec:reference_systems}
\begin{figure*}
    \includegraphics[width=\textwidth]{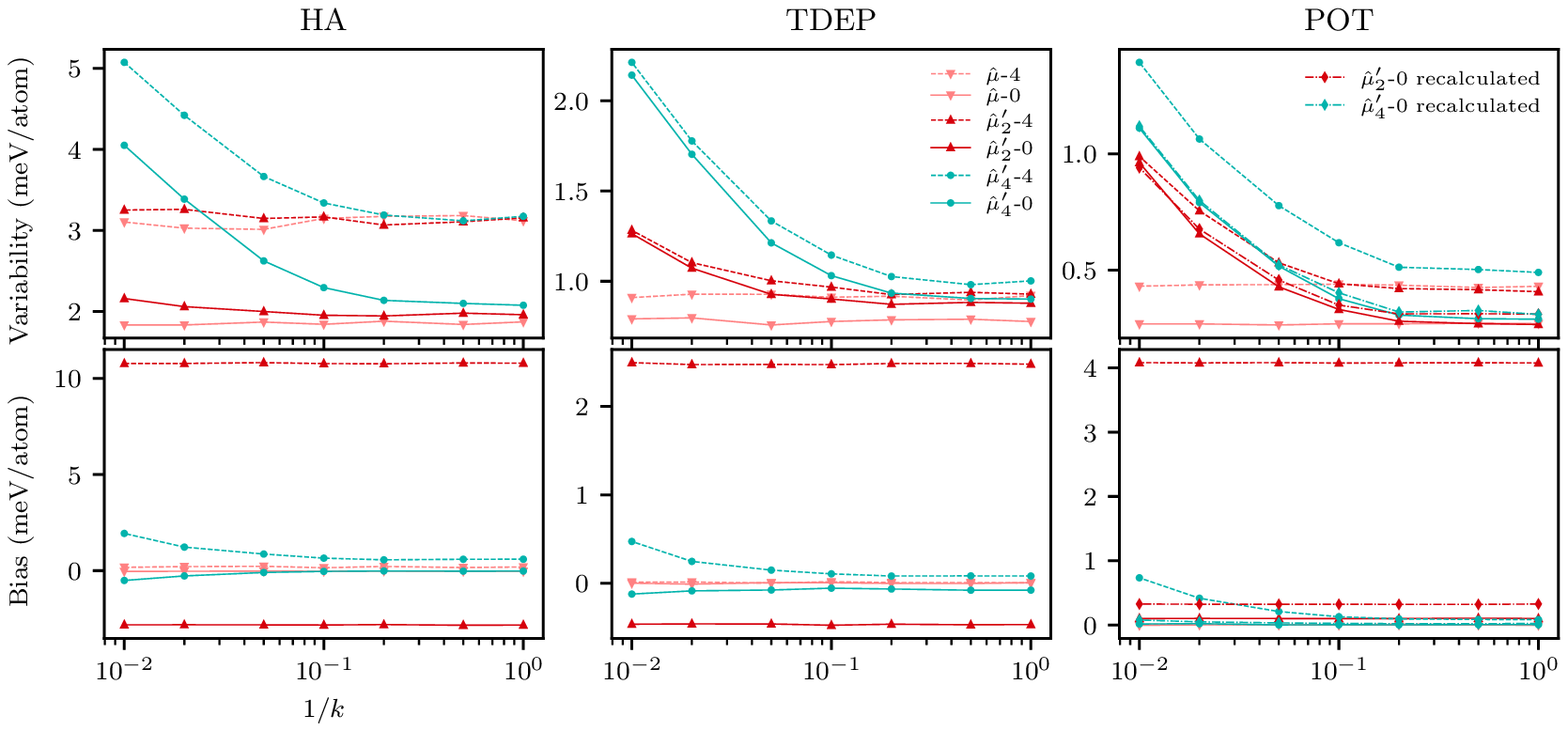} 
    \caption{\label{fig:111_vs_444} Dependence of the variabilities and 
    biases of different $\Delta F$ estimators on various reference 
    potentials and the fraction of recalculated structures for a 
    simulation time of \SI{10}{\pico\second} and \mbox{SNAP-4} as the 
    potential $\ulo$. The variability is expressed as half of the range 
    that contains 95\% of the sampled free energies. Since the sampling 
    distributions were approximately normal, this is close to $1.96 
    \sigma(\dhatf)$. The bias is defined as $E[\dhatf]-\Delta F$, where $\Delta F$ was estimated using all of the 
    available data and the estimator $\muhat$. The dashed and solid 
    lines denote that \mbox{SNAP-4} and \mbox{SNAP-0} were used 
    respectively to obtain the reference potential. The dashed and 
    dotted lines indicate that the fitting was done using  forces and 
    energies recalculated using \mbox{SNAP-0} of the configurations 
    sampled by \mbox{SNAP-4}. 
    }
\end{figure*}

It was qualitatively explained in Section~\ref{sec:theory} how the 
choice of the reference system could affect the accuracy of the 
results, but without knowing the specific potentials involved, it was 
not possible to predict how much difference does it make in practice. 
The results of using either HA, TDEP or POT as $\uharm$ and {SNAP-4} as 
$\ulo$ are given in Figure~\ref{fig:111_vs_444}. 

As expected, in terms of variability, POT is the best reference, 
followed by TDEP and HA. Using the former makes it possible to 
determine both $\Delta F(\muhattwo)$ and $\Delta F(\muhatfour)$ to a 
precision of less than 1~meV/atom with a relatively short simulation 
time of \SI{10}{\pico\second} and only about 10 to 20 recalculated 
energies at each $\lambda$. With the other two references the variance 
of $\ulo-\uharm$ increases the baseline of the uncertainty 
significantly so either more recalculations, longer simulations, or 
both are needed to achieve similar precision. In each case the variance 
of $\dftwo$ is smaller than that of $\dffour$ meaning that for a given target precision, the former is 
a more efficient estimator, however the difference becomes smaller with 
a better reference system.

\begin{figure}
    \includegraphics[width=8cm]{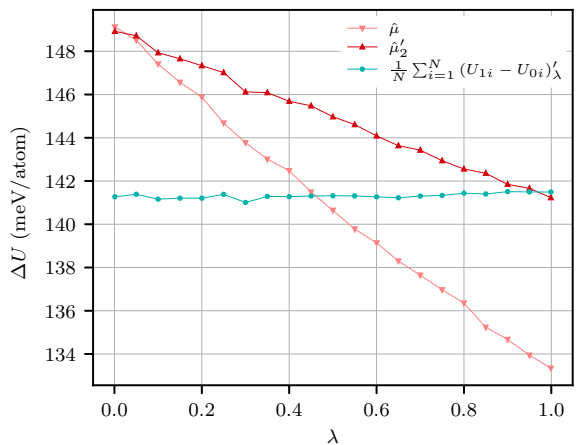}
    \caption{\label{fig:uptild_bias} The switching curves from
    POT fitted to SNAP-4 at $\lambda=0$ to SNAP-4
    at $\lambda=1$. Each point
    is averaged over a \SI{60}{\pico\second} simulation. In $\muhattwo$
    and the correction term all of the energies were recalculated,
    \ie $k = 1$ in Equation~\ref{eq:muhatuptild}.
    }
\end{figure}

The situation becomes different and more complex when the bias is also 
taken into account. Firstly, in every case the bias of $\dffour$ is 
smaller than that of $\dftwo$ and at least for the small system 
considered also significantly smaller than the variability. 
Interestingly, the bias of $\dftwo$ can be excessive even when using a 
good reference system. In addition, it is not necessarily easy to 
quantify without directly comparing $\muhattwo$ and $\muhat$. For 
example, it has been proposed that one of the measures for the 
applicability of the UP-TILD method is that the correction term 
$\frac{k}{N} \sum_{i=1}^{N/k} \left(\uhi{_{ki}}-\U\lo{{ki}}\right)$ 
would be nearly independent of $\lambda$ \cite{Grabowski2009}. As shown 
in Figure~\ref{fig:uptild_bias}, this is not always the case. With POT 
fitted to {SNAP-4} as $\uharm$ this term varies by much less than 
1~meV/atom, whereas the bias of $\muhattwo$ grows linearly with 
$\lambda$ and is over 4~meV/atom for $\dhatf$. Conversely, with 
POT fitted to SNAP-0 as $\uharm$, the correction term changes by 
7.9~meV/atom between $\lambda=0$ and $\lambda=1$, while the bias is 
negligible. In short, the estimator of the bias based on the proposal
above could itself be biased.

\begin{figure*}
    \includegraphics[width=\textwidth]{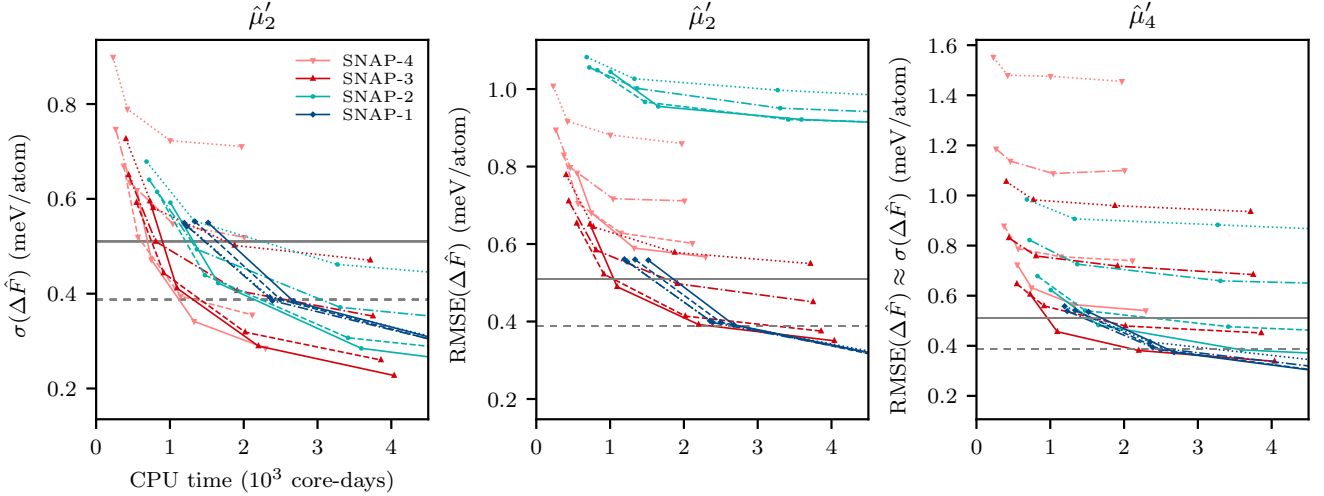} 
    \caption{\label{fig:performance} Comparison of the performance of 
    using $\muhattwo$ and $\muhatfour$ for estimating $\Delta F$. The 
    points in increasing order of the CPU time on each line correspond 
    to 10, 20, 50 and 100 ps simulation times at each of the 11 
    equidistant $\lambda$ values. The dotted, dashed-dotted, dashed and 
    solid lines denote 10, 20, 50 and 100 recalculations at each 
    $\lambda$ respectively. Horizontal solid and dashed are at 1/1.96 
    and 1/2.58 meV/atom, \ie the values at which in the absence of bias 
    95\% and 99\% of the results would be within 1 meV of the true 
    $\Delta F$. For reference, the corresponding CPU times when using 
    \mbox{SNAP-0} and $\muhat$ would be $11\cdot10^3$ and $20\cdot10^3$ 
    core-hours. $\sigma$ denotes the standard deviation and 
    $\mathrm{RMSE}^2 = E[(\Delta\hat{F}-\Delta F)^2]$. The CPU time for 
    each $\ulo$ is the estimated computational time if the 
    corresponding approximation of DFT was used.
    }
\end{figure*}
% 21 lambdas SNAP-0 21.6457765 37.52319142

Given that the bias of $\dffour$ is typically significantly
smaller than that of $\dftwo$, then the latter can also be approximated as 
\begin{equation}
\begin{aligned}
E[\dftwo]-\Delta F &\approx E[\dftwo]-E[\dffour] \\
&= \int_0^1 E \left[ \muhattwo-\muhatfour \right] d\lambda 
\end{aligned}
\end{equation}

Since in practice the expectation values in the equation above will be 
replaced by a single sample, the estimate of the bias can have large 
uncertainty for large values of $k$. This can make it less useful in 
the cases where $\var [\dffour ]$ is significantly larger than $\var 
[\dftwo]$, however, it is asymptotically correct.

As evident, there can be a significant difference in the results 
whether the reference potential is fitted to $\uhi$ or $\ulo$. In every 
case the former is a better choice with the largest change observed in 
the bias of $\dftwo$. There is also a great improvement in both the 
variance and the bias when the reference potential is fitted indirectly 
to $\uhi$ by first sampling the configurations using $\ulo$ and then 
recalculating the energies and forces of those using $\uhi$ to be used 
in the fitting database. Whereas those configurations are not 
necessarily the same as those from direct sampling, it can be expected 
that the fitted potential is at least somewhat transferable, therefore 
providing an improvement.

\begin{figure}
    \includegraphics[width=8cm]{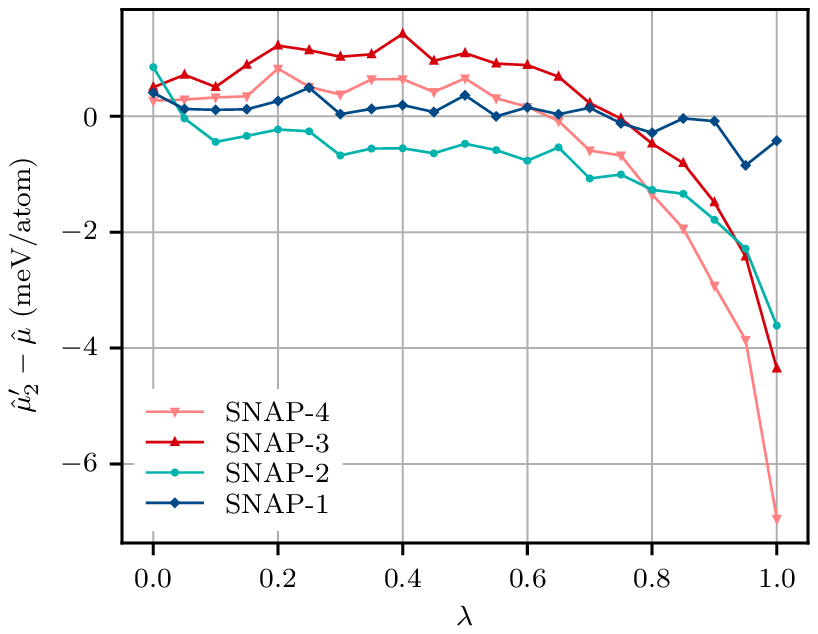} 
    \caption{\label{fig:uptild_biases} Estimates of the biases of 
    $\muhattwo(\lambda)$ for different $\ulo$. The integrals of 
    the curves are the biases of $\dhatf (\muhattwo)$.
    }
\end{figure}

\subsection{Choice of the approximate Hamiltonian}

\begin{figure*}
    \includegraphics{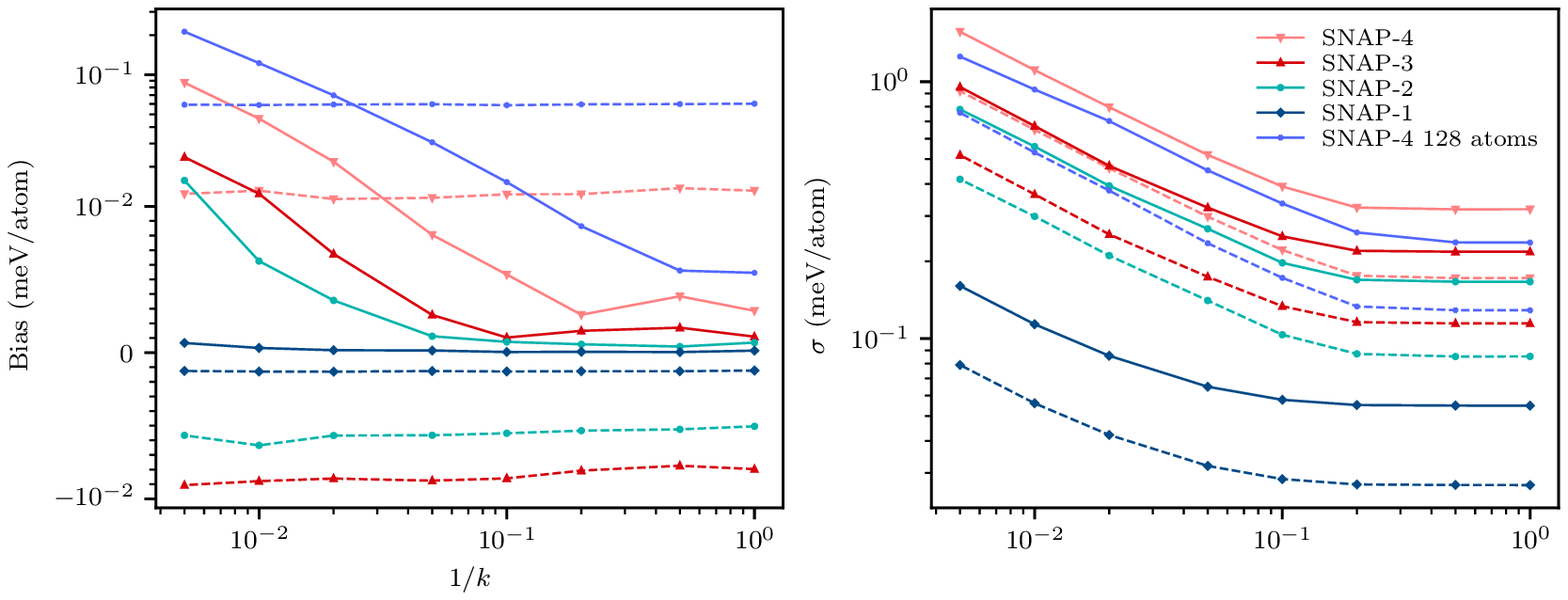}\label{fig:fep_bias_std}
    \caption{\label{fig:fep_bias_std} The bias and standard deviation
    of $F_1 - F_1'$ estimated by free energy perturbation from
    a \SI{50}{\pico\second} simulation. The solid and dashed line
    denote the exact (Equation~\ref{eq:hatfep}) or the approximate
    (Equation~\ref{eq:hatfepvar}) equation was used. Note that
    the scale in the bias plot is linear between $-10^{-2}$ and $10^{2}$ meV/atom and logarithmic
    otherwise.
    }
\end{figure*}

Choosing which DFT parameters to modify in order to speed up the 
calculations is not obvious. For the sake of argument, suppose that 
using Equation~\ref{eq:muhatuptild} does not result in excessive bias 
so it can be used to obtain good estimates of $\Delta F$ and in addition the 
covariance between the first and the second term is zero, in which case 
the total variance is the sum of the variances of the two terms. If 
$\ulo$ is a close approximation of $\uhi$, then by definition 
${\operatorname{var}(\ulo-\uhi)}$ is small and not many recalculations 
are needed to have the uncertainty of the second sum be sufficiently 
small. At the same time the MD simulation itself is slower and 
therefore the time to statistically convergence the first term longer 
compared to using a worse approximation. In the latter case 
${\operatorname{var}(\ulo-\uhi)}$ is larger, so either more 
recalculations are needed to converge the second sum to the same level 
as with a better $\ulo$ or the simulation needs to be run longer in 
order to get a better convergence of the first sum such that the total 
uncertainty remains the same. If the time to perform the additional 
calculations are compensated by the faster speed of the worse 
approximation, the total computational time is reduced.

Given the reasoning above, it is probably not possible to give 
universal guidelines for choosing optimal $\ulo$, since it depends on 
the chosen type of $\muhat$, atomic system, its size, reference and 
target potentials, required accuracy and available computational 
resources. Therefore the results presented here should be taken as one 
illustration of many possible outcomes.

A comparison between the computational times of $\dftwo$ and $\dffour$ 
with different approximating potentials $\ulo$ and TDEP of {SNAP-0} as 
$\uharm$ is given in Figure~\ref{fig:performance}. The substantial 
overlap of different lines on each plot, especially on the first one 
depicting the standard deviation of $\dftwo$, is a good example of what 
was described above. This indicates that for a given error and 
computational time there can be several equivalent solutions in terms 
of chosen $\ulo$, the simulation time and the number of recalculations. 
For example, it is seen that there is only a slight difference in 
$\sigma(\dhatf)$ at around 1000 core-hours of total 
computational time whether 20 recalculations of 50~ps simulations, 100 
recalculations of 10~ps simulations or 10 recalculations of 10~ps 
simulations at each $\lambda$ are done using \mbox{SNAP-4}, 
\mbox{SNAP-2} and \mbox{SNAP-1} respectively as $\ulo$. 

It is also clear that recalculating more energies lowers the standard 
deviation significantly when $\ulo$ is a bad approximation of $\uhi$, as 
is the case with {SNAP-4}, and that there is almost no change except for 
added computational time with a good approximation, such as {SNAP-1}. 
In addition, it is seen that although {DFT-4} is about 6 times faster 
than {DFT-1} in MD simulations, it is at best only about 2 times as 
fast when approximating $\Delta F$ due to the high computational
cost of recalculations.

There is a significant change in the results when the bias of $\dftwo$ 
is taken account in the error, as shown in the second plot of 
Figure~\ref{fig:performance}. The biases of $\dftwo$ are $0.05$, 
$-0.8$, $0.3$, $-0.5$ and meV/atom for {SNAP-1} to {SNAP-4} 
respectively. It is notable that apart from {SNAP-1}, the bias seems 
essentially random. Whereas it could be expected that {SNAP-2} is a 
significantly better approximation of {SNAP-0} than {SNAP-4}, its bias 
is considerably larger. As shown in Figure~\ref{fig:uptild_biases}, 
this is due to different cancellation of the biases of $\muhattwo$ when 
integrating over $\lambda$. If instead the error would be defined as 
$\sqrt{ \int_0^1 (\muhattwo-\muhat)^2 d\lambda}$, the corresponding 
values would be $0.2$, $1.1$, $1.3$ and $1.7$ meV/atom.

Whether the bias of $\dftwo$ can be considered small enough depends on 
the application. When free energy differences between different phases 
are compared, it is possible that the biases either cancel out or add 
and as shown, the sign of each can depend on the chosen $\ulo$. 
Furthermore, if the bias is highly non-linear with temperature or 
volume, it can also have a considerable effect on the derivative 
properties, such as the heat capacity or thermal expansion coefficient. 
Therefore in order to obtain accurate results it is best not to rely on 
the possible cancellation effect which would mean constraining the use 
of $\dftwo$ to only cases when $\ulo$ is a good approximation of 
$\uhi$.

\begin{figure*}
    \includegraphics[width=\textwidth]{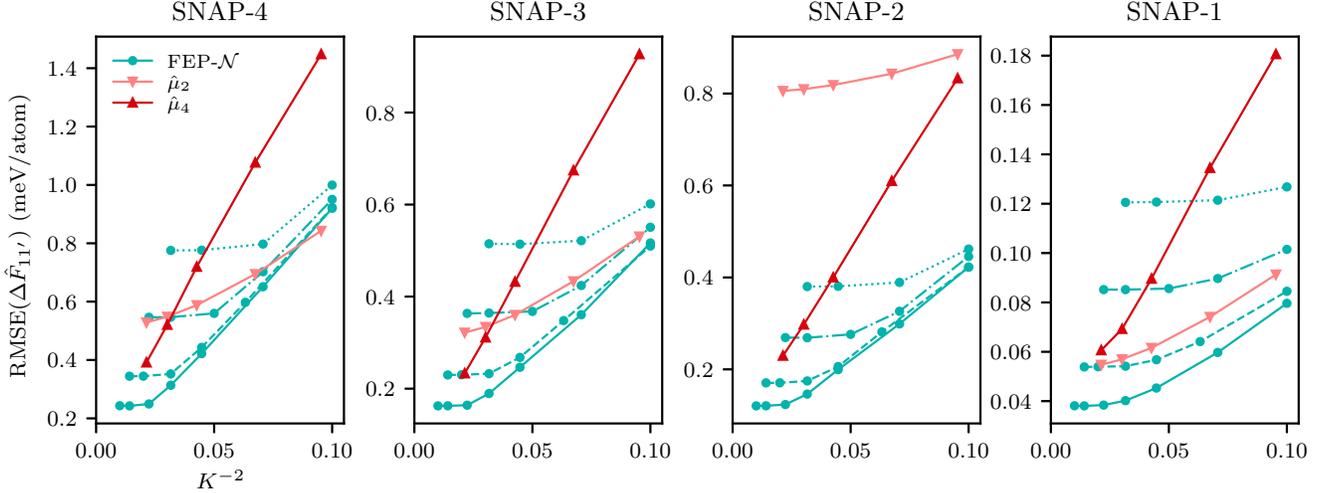} 
    \caption{\label{fig:fep_comparison} Comparison of the RMS error
    of the different estimators of $F_1-F_1'$ using either cumulant expansion FEP or $\lambda$-switching.
    In each case the reference potential $\uharm$ was TDEP fitted to $\uhi$ and the approximating potential
    $\ulo$ is given in the title of each subplot.
    The dotted, dashed-dotted, dashed and solid lines denote simulation times
    of 10, 20, 50 and 100~ps respectively. $K$ denotes the total number of recalculated
    structures which in case of $\muhattwo$ and $\muhatfour$ based estimators was
    divided evenly between 11 equidistant $\lambda$ values.}
\end{figure*}

As explained before, another way to avoid biased results is to use 
$\dffour$. For the results shown in Figure~\ref{fig:performance}, the 
bias was typically less than 0.1 meV/atom, therefore $\sigma(\hat{F})$ 
was very close to $\mathrm{RMSE}(\hat{F})$. The reduced bias comes, 
however, with the cost of increased variance with an exception when 
\mbox{SNAP-1} is used as $\ulo$ as in that case there is essentially no 
difference between $\dftwo$ and $\dffour$. This is due to 
$\operatorname{var} (\uhi-\ulo)$ being small enough that essentially 
for any number of recalculations the error is dominated by 
$\operatorname{var} (\ulo-\uharm)$.

%\includegraphics[width=\textwidth]{std_vs_pot_multi_ver3.eps}
    %\caption{\label{fig:std_vs_pot} uncert vs pot}
    
\subsection{Comparison to FEP}
As explained in Section~\ref{sec:theory}, the correction term

\begin{equation}\label{eq:mucorr}
\hat{\mu}'_c(\lambda) = \int_0^1 \left[ \muhatprime (\lambda) - \frac{1}{N} \displaystyle\sum_{i=1}^{N} \left(\U\lo{i}-\uharm{_i}\right)_\lambdalo \right] d\lambda
\end{equation}

with $\muhatprime$ being either $\muhattwo$ of $\muhatfour$, estimates 
the free energy difference $\Delta F_{11'} = F_1 - F_1'$. Since both of 
the estimators depend on the reference potential $\uharm$ (it does not 
appear explicitly in the correction term when using $\muhattwo$, but is 
included in the potential $\U'$ that is used for sampling), so do the 
bias and variance of the estimated $\Delta F_{11'}$ which in turn 
propagates to the error of $\dhatf$.

Free energy perturbation provides a more natural way of estimating 
$\Delta F_{11'}$ such that its error does not depend on $\uharm$, but 
only on the distribution of $\uhi-\ulo$ sampled at $\lambda=1$. As the 
results in Figure~\ref{fig:fep_bias_std} show, when using either 
$\deltahatffep$ or $\deltahatffepn$ the uncertainty of the results in 
all cases is dominated by the standard deviation instead of bias with 
the latter being approximately an order of magnitude smaller. This can 
be attributed mainly to the small system size which limits $\var 
[\beta(\uhi-\ulo)]$ to reasonably small values (1.4 with 54 and 3.3 
with 128 atoms) even for the worst $\ulo$ . The bias of 
$\deltahatffepn$ does not depend significantly on the number of 
recalculated energies, which is expected since the sample variance is an 
unbiased estimator of the population variance. The value of the bias, 
although not zero, which indicates that the distribution of $\uhi-\ulo$ 
is not perfectly Gaussian, is nevertheless small enough that it can be 
considered negligible and $\deltahatffepn$ is therefore a good 
approximation for the potentials used in this work. The main advantage 
of $\deltahatffepn$ compared to $\deltahatffep$ is the approximately 
two times decrease in the standard deviation of the estimated free 
energy difference. Given uncorrelated samples, about four times fewer 
calculations would be needed to reach the same uncertainty, however, 
the improvement can be even higher as evident by the leveling off of 
the decrease in $\sigma$ when more than 10-20\% energies are 
recalculated.

The latter also means that given some fixed number of samples at each 
$\lambda$ that is needed to converge the uncorrected free energy 
difference $F_1'-F_0$ (Equation~\ref{eq:df_uncorrected}), it is 
possible that using FEP might not be as efficient for estimating $F_1-F_1'$ 
compared to Equation~\ref{eq:mucorr}. Assuming that each recalculated 
energy provides certain constant amount of information about 
$F_1-F_1'$, then instead of recalculating, for example, 5\% of energies 
at 10 values of $\lambda$ when calculating $\dftwo$ or $\dffour$, with 
FEP $10 \cdot 5 = 50\%$ energies would have to be recalculated in order 
to get the same uncertainty. If that fraction of energies is greater 
than the threshold above which the correlations between samples become 
large enough, there is no reduction in the overall uncertainty. 

In practice, the comparison is more complicated mainly because each 
recalculated sample does not necessarily provide the same amount of 
information. It can expected that with $\dftwo$ and $\dffour$ either 
recalculating different number of energies at each $\lambda$ or 
assigning different weights to each $\muhat$ when fitting the 
polynomial could result in slightly better $\sigma(\dhatf)$. 
Whereas this was not as crucial for the results in the previous 
sections since the comparison was between methods that used the same 
data, this is not the case when estimating $\Delta F_{11'}$ using FEP.
Consequently a full analysis would deserve a separate study and the 
results presented here are only for the case when neither the length
of the simulation nor the number of recalculated energies do not depend
on $\lambda$ and the comparison will only be made for estimating $F_1-F_1'$,
not $F_1-F_0$. 

The results in Figure~\ref{fig:fep_comparison} illustrate what was 
pointed out above. \FEPN can have larger error when estimating 
$F_1-F_1'$ compared to $\muhattwo$ or $\muhatfour$ for short simulation 
times as the decrease in variance due to added number of recalculations 
will be limited due to time correlations. For longer simulation times, 
in which case both $\muhat$-based estimators and \FEPN make use of only 
uncorrelated data, the latter is more efficient and has smaller error. 
In other words, at least for the given potentials and system size, 
\FEPN makes better use of the recalculated energies given that those 
are randomly sampled. FEP, however, is not necessarily better due to
significantly larger variance.

\subsection{Comparisons of $\muhatone$ to $\muhattwo$ and $\muhatthree$ to $\muhatfour$}\label{sec:muhat_comparison}
\begin{figure}
    \includegraphics[width=\columnwidth]{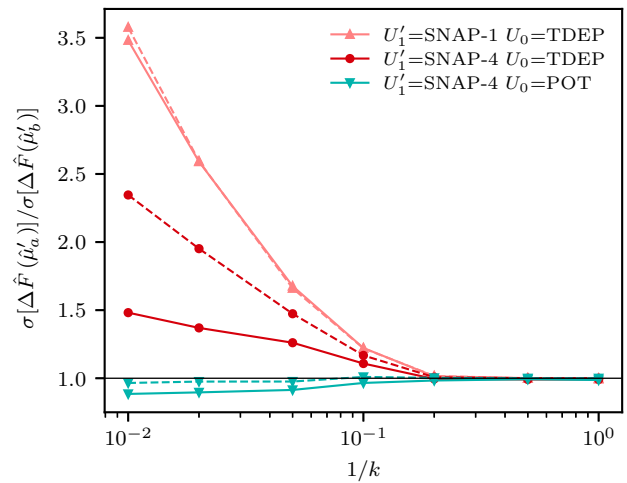} 
    \caption{\label{fig:muhat3vs4} 
        Ratio of the standard deviations of the free energy estimators 
        based on either all of the available data or only those of  the 
        recalculated timesteps. The solid lines correspond to $a=3$, 
        $b=4$ and the dashed lines to $a=1$, $b=2$. Values below 1 
        indicate the cases when using the free energy estimator based 
        on fewer data results in lower standard deviation. $1/k$ is the 
        fraction of the number of recalculated energies. }
\end{figure}

The estimators of $\Delta F$ presented so far have been based on either 
$\muhattwo$ and $\muhatfour$ and not $\muhatone$ or $\muhatthree$, \ie the 
variants that only incorporate energy differences corresponding to the 
recalculated timesteps. It would be natural to expect that using fewer 
samples would result in increased standard deviation, which is the case 
for a regular sample mean, but as shown in Figure~\ref{fig:muhat3vs4} 
this is not necessarily the case. The results show that among the 
combinations of $\ulo$ and $\uharm$ considered in this study, the 
standard deviation can indeed increase significantly, but also decrease. 
Because those cases are not in any way guaranteed to form a 
representative sample and do not explain the magnitude and the 
direction of the change in the standard deviation, it is worthwhile to 
investigate it theoretically. For simplicity, only the comparison
between $\muhatone$ and $\muhattwo$ is given.

If three sample means are defined as

\begin{subequations}
 \label{eq:sums}
 
 \begingroup
%\addtolength{\jot}{1.0em}
\begin{align}
\hat{m}_1 &= \frac{k}{N} \sum_{i=1}^{N/k} \left(\U\lo{{ki}}-\U\harm{_{ki}}\right)_\lambdalo \\
\hat{m}_2 &= \frac{k}{N} \sum_{i=1}^{N/k} \left(\uhi{_{ki}}-\U\lo{{ki}}\right)_\lambdalo \\
\hat{m}_3 &= \frac{1}{N} \sum_{i=1}^{N} \left(\U\lo{{i}}-\U\harm{_{i}}\right)_\lambdalo \label{eq:sum3}
\end{align}
\endgroup
\end{subequations}

then $\muhatone$ and $\muhattwo$ can be written as

\begin{subequations}

 \begingroup
%\addtolength{\jot}{1.0em}
\begin{align}
\muhatone &= \hat{m}_1 + \hat{m}_2 \\
\muhattwo &= \hat{m}_3 + \hat{m}_2
\end{align}
\endgroup
\end{subequations}

Although the $N$ samples in Equation~\ref{eq:sum3} are usually 
correlated, assuming that they are random does not qualitatively change 
the results and in that case

\begin{equation} \label{eq:var_ratio_k}
\var(\hat{m}_1) = k \var(\hat{m}_3) 
\end{equation}

Then by defining $\gamma$ and $\rho$ as

\begin{subequations}
 \begingroup
\begin{align}
\var (\hat{m}_1) &= \gamma \var(\hat{m}_2) \\
\rho &= \operatorname{corr}(\hat{m}_1,\hat{m}_2)
\end{align}
\endgroup
\end{subequations}

it can be shown that

\begin{equation}
\frac{\var(\muhatone)}{\var(\muhattwo)} = \eta(\rho,\gamma,k) =  \frac{\gamma + 2 \rho \sqrt{\gamma} +1 }{\gamma/k + 2 \rho \sqrt{\gamma}/k +1 }
\end{equation}

\begin{figure}
    \includegraphics[width=\columnwidth]{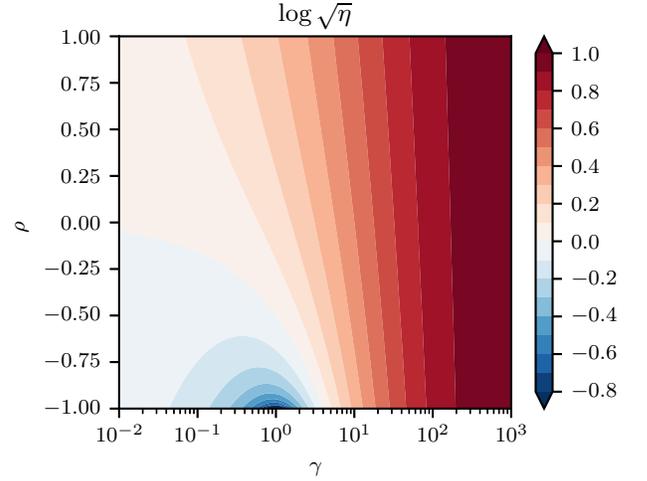} 
    \caption{\label{fig:std_ratios} Dependence of $\eta$ on $\gamma$ 
    and $\rho$ at $k=100$. Region where $\eta < 1$ indicates conditions 
    when using fewer data, \ie $\muhatone$ instead of $\muhattwo$,
    results in smaller variance.}
\end{figure}

The values of this function at fixed $k=100$ are shown in 
Figure~\ref{fig:std_ratios}. Irrespective of the value of $k$, the 
condition for $\eta < 1$, \ie that using fewer samples results in
smaller variance, is

\begin{equation}\label{eq:eta1border}
\rho \le -\frac{\sqrt{\gamma}}{2}
\end{equation}

Which means that in addition to $\rho$ being negative the value of $\gamma$ 
can be at most 4.

Since

\begin{subequations}
\begingroup
\begin{align}
\var(\hat{m}_1) &= k/N \var \left(\uharm -\ulo\right) \\
\var(\hat{m}_2) &= k/N \var \left(\uhi -\ulo \right)
\end{align}
\endgroup
\end{subequations}

then $\gamma$ can also be written as

\begin{equation}
\gamma = \frac{\var \left(\uharm -\ulo\right)}{\var \left(\uhi -\ulo \right)}
\end{equation}

and using similar reasoning

\begin{equation}
\rho = \operatorname{corr}(\ulo -\uharm, \uhi -\ulo)
\end{equation}

This suffices to explain the results in Figure~\ref{fig:muhat3vs4}. 
Since typically $\uharm$ does not approximate $\ulo$ as well as $\uhi$, 
\mbox{$\gamma \gg 1$} and it is better to use $\muhattwo$ over $\muhatone$. 
This is observed for both cases when TDEP was used as the reference 
potential. With {SNAP-1} as $\ulo$, $\eta$ is larger due to the 
variance of $\uhi -\ulo$ being significantly smaller. At the limit of 
infinite $\gamma$, $\eta$ approaches $k$. 

At $\gamma =1$ and $\rho = -1$, for example when $\uharm = \uhi$, 
$\eta$ becomes zero. SNAP as a reference potential, although not 
perfect, is able to get closer to that point compared to TDEP, in this 
case on average over all $\lambda$ values $\gamma = 1.8$ and $\rho = 
-0.71$. This satisfies Equation~\ref{eq:eta1border} which results in 
$\dhatf (\muhatone)$ having slightly lower variance than 
$\dhatf (\muhattwo)$. Whereas not investigated here, it must
be noted that as $\uharm$ approaches $\uhi$, the need for using $\ulo$
disappears, since it likely becomes better to estimate $F_1-F_0$ directly
by FEP or \FEPN.

% avg over all lambdas
% 222_223 tdep rho = 0.12, gamma = 10^2.8
% 111_low tdep rho = -0.30, gamma = 10^1.0
% 111_low snap rho = -0.71, gamma = 10^0.25

\section{Conclusions}
Even though only a single atomic system was investigated and the 
numerical calculations were performed using SNA potentials fitted to 
DFT instead of the latter directly, the results nevertheless provide 
insight into the possible issues and pitfalls when employing different 
correction schemes for estimating converged free energy differences 
from non-converged DFT-MD simulations.

The choice between estimating the free energy difference between a 
target and a reference system either by using the UP-TILD method 
($\dftwo$) or by weighted ensemble averages ($\dffour$) determines 
whether the error of the results is dominated by bias, as is the case 
for the former method, or by variance as is the case for the latter. As 
opposed to $\dffour$, the bias of $\dftwo$ cannot be reduced by longer 
simulations or by recalculating a larger number of energies using the 
converged potential $\uhi$. This limits the choice of both the 
reference ($\uharm$) and the approximating ($\ulo$) potentials, which 
have to be relatively close approximations of $\uhi$. Most importantly, 
fitting $\uharm$ to $\ulo$ instead of $\uhi$ can significantly increase 
the bias, in some cases by more than an order of magnitude, and should 
therefore be avoided. In addition, an anharmonic $\uharm$ is 
substantially better than a harmonic one both for reducing the bias and 
the variance of either estimator. Whereas this complicates the 
analysis, since the free energy of the reference system has to be 
calculated separately, for a given computational time it allows for 
much more accurate and precise results.

As opposed to $\dftwo$, $\dffour$ could estimate $\Delta F$ accurately 
even using the worst $\ulo$ and $\uharm$. For the combinations of 
potentials, simulation times and the number of recalculated energies 
considered in this paper, the bias of $\dffour$ did not pose a problem 
since the variance of it was significantly larger. Therefore, by 
minimizing the variance to an acceptable level, the bias was reduced to 
be insignificant. However, the variance was in every case larger than that 
of $\dftwo$, which requires longer simulations and more recalculations 
in order to achieve the same precision.

Choosing an optimal $\ulo$ can be difficult. Using a bad approximation 
results in faster MD simulations, but due to the increased variance 
more recalculated energies are needed and for a given target precision 
the total computational time can be longer compared to when a better 
$\ulo$ is used. Since poor $\ulo$ can also significantly increase the 
bias of $\dftwo$ or the variance of $\dffour$, although possibly not 
the optimal one, a relatively safe choice is DFT with only a slightly 
reduced convergence parameters.

Both of the correction schemes estimate $F_1-F_1'$, \ie the free energy 
difference between systems with converged and non-converged potentials. 
As opposed to the direct thermodynamic path, this is done on a path 
between the non-converged potential to the reference. Therefore the 
estimated $F_1-F_1'$ depends on $\uharm$ whereas the actual free energy 
difference does not. This can reduce the efficiency and accuracy 
compared to a direct FEP estimate. However, as the results show, FEP is 
not necessarily better in every case. The main reason behind this is 
that distributing the recalculated energies among all the 
$\lambda$-values can result in a larger amount of uncorrelated samples 
than doing it at only $\lambda=1$, as is the case with FEP. On the 
other hand, even with the worst $\ulo$, the bias did not pose a problem 
neither with FEP nor its cumulant expansion approximation. 

Finally, for both $\dftwo$ and $\dffour$ there are corresponding 
estimators that as opposed to all of the MD data, only use that which 
corresponds to the recalculated timesteps. Whereas in general using 
more data results in smaller variance, in some cases, namely when using 
a good reference potential, the opposite can be true.

\section{Acknowledgments}
The research leading to these results has been partially funded by the 
Swedish Centre for Nuclear Technology (SKC). Computational resources 
were provided by Swedish National Infrastructure for Computing (SNIC).

\bibliography{uptild,vasp_cite}{}

%Control: production of eprint (1) enabled
\begin{thebibliography}{47}%
\makeatletter
\providecommand \@ifxundefined [1]{%
 \@ifx{#1\undefined}
}%
\providecommand \@ifnum [1]{%
 \ifnum #1\expandafter \@firstoftwo
 \else \expandafter \@secondoftwo
 \fi
}%
\providecommand \@ifx [1]{%
 \ifx #1\expandafter \@firstoftwo
 \else \expandafter \@secondoftwo
 \fi
}%
\providecommand \natexlab [1]{#1}%
\providecommand \enquote  [1]{``#1''}%
\providecommand \bibnamefont  [1]{#1}%
\providecommand \bibfnamefont [1]{#1}%
\providecommand \citenamefont [1]{#1}%
\providecommand \href@noop [0]{\@secondoftwo}%
\providecommand \href [0]{\begingroup \@sanitize@url \@href}%
\providecommand \@href[1]{\@@startlink{#1}\@@href}%
\providecommand \@@href[1]{\endgroup#1\@@endlink}%
\providecommand \@sanitize@url [0]{\catcode `\\12\catcode `\$12\catcode
  `\&12\catcode `\#12\catcode `\^12\catcode `\_12\catcode `\%12\relax}%
\providecommand \@@startlink[1]{}%
\providecommand \@@endlink[0]{}%
\providecommand \url  [0]{\begingroup\@sanitize@url \@url }%
\providecommand \@url [1]{\endgroup\@href {#1}{\urlprefix }}%
\providecommand \urlprefix  [0]{URL }%
\providecommand \Eprint [0]{\href }%
\providecommand \doibase [0]{https://doi.org/}%
\providecommand \selectlanguage [0]{\@gobble}%
\providecommand \bibinfo  [0]{\@secondoftwo}%
\providecommand \bibfield  [0]{\@secondoftwo}%
\providecommand \translation [1]{[#1]}%
\providecommand \BibitemOpen [0]{}%
\providecommand \bibitemStop [0]{}%
\providecommand \bibitemNoStop [0]{.\EOS\space}%
\providecommand \EOS [0]{\spacefactor3000\relax}%
\providecommand \BibitemShut  [1]{\csname bibitem#1\endcsname}%
\let\auto@bib@innerbib\@empty
%</preamble>
\bibitem [{\citenamefont {Bendick}\ and\ \citenamefont
  {Pepperhoff}(1982)}]{Bendick1982}%
  \BibitemOpen
  \bibfield  {author} {\bibinfo {author} {\bibfnamefont {W.}~\bibnamefont
  {Bendick}}\ and\ \bibinfo {author} {\bibfnamefont {W.}~\bibnamefont
  {Pepperhoff}},\ }\bibfield  {title} {\bibinfo {title} {{On the
  $\alpha$/$\gamma$ phase stability of iron}},\ }\href
  {https://doi.org/10.1016/0001-6160(82)90117-1} {\bibfield  {journal}
  {\bibinfo  {journal} {Acta Metallurgica}\ }\textbf {\bibinfo {volume} {30}},\
  \bibinfo {pages} {679} (\bibinfo {year} {1982})}\BibitemShut {NoStop}%
\bibitem [{\citenamefont {Dinsdale}(1991)}]{Dinsdale1991}%
  \BibitemOpen
  \bibfield  {author} {\bibinfo {author} {\bibfnamefont {A.~T.}\ \bibnamefont
  {Dinsdale}},\ }\bibfield  {title} {\bibinfo {title} {{SGTE data for pure
  elements}},\ }\href {https://doi.org/10.1016/0364-5916(91)90030-N} {\bibfield
   {journal} {\bibinfo  {journal} {Calphad}\ }\textbf {\bibinfo {volume}
  {15}},\ \bibinfo {pages} {317} (\bibinfo {year} {1991})}\BibitemShut
  {NoStop}%
\bibitem [{\citenamefont {Grabowski}\ \emph {et~al.}(2011)\citenamefont
  {Grabowski}, \citenamefont {S{\"{o}}derlind}, \citenamefont {Hickel},\ and\
  \citenamefont {Neugebauer}}]{Grabowski2011a}%
  \BibitemOpen
  \bibfield  {author} {\bibinfo {author} {\bibfnamefont {B.}~\bibnamefont
  {Grabowski}}, \bibinfo {author} {\bibfnamefont {P.}~\bibnamefont
  {S{\"{o}}derlind}}, \bibinfo {author} {\bibfnamefont {T.}~\bibnamefont
  {Hickel}},\ and\ \bibinfo {author} {\bibfnamefont {J.}~\bibnamefont
  {Neugebauer}},\ }\bibfield  {title} {\bibinfo {title} {{Temperature-driven
  phase transitions from first principles including all relevant excitations:
  The fcc-to-bcc transition in Ca}},\ }\href
  {https://doi.org/10.1103/PhysRevB.84.214107} {\bibfield  {journal} {\bibinfo
  {journal} {Physical Review B - Condensed Matter and Materials Physics}\
  }\textbf {\bibinfo {volume} {84}},\ \bibinfo {pages} {214107} (\bibinfo
  {year} {2011})}\BibitemShut {NoStop}%
\bibitem [{\citenamefont {Burke}(2012)}]{Burke2012}%
  \BibitemOpen
  \bibfield  {author} {\bibinfo {author} {\bibfnamefont {K.}~\bibnamefont
  {Burke}},\ }\bibfield  {title} {\bibinfo {title} {{Perspective on density
  functional theory}},\ }\href {https://doi.org/10.1063/1.4704546} {\bibfield
  {journal} {\bibinfo  {journal} {Journal of Chemical Physics}\ }\textbf
  {\bibinfo {volume} {136}},\ \bibinfo {pages} {150901} (\bibinfo {year}
  {2012})}\BibitemShut {NoStop}%
\bibitem [{\citenamefont {Jones}(2015)}]{Jones2015}%
  \BibitemOpen
  \bibfield  {author} {\bibinfo {author} {\bibfnamefont {R.~O.}\ \bibnamefont
  {Jones}},\ }\bibfield  {title} {\bibinfo {title} {{Density functional theory:
  Its origins, rise to prominence, and future}},\ }\bibfield  {journal}
  {\bibinfo  {journal} {Reviews of Modern Physics}\ }\textbf {\bibinfo {volume}
  {87}},\ \href {https://doi.org/10.1103/RevModPhys.87.897}
  {10.1103/RevModPhys.87.897} (\bibinfo {year} {2015})\BibitemShut {NoStop}%
\bibitem [{\citenamefont {Palumbo}\ \emph {et~al.}(2014)\citenamefont
  {Palumbo}, \citenamefont {Burton}, \citenamefont {Costa~e silva},
  \citenamefont {Fultz}, \citenamefont {Grabowski}, \citenamefont {Grimvall},
  \citenamefont {Hallstedt}, \citenamefont {Hellman}, \citenamefont {Lindahl},
  \citenamefont {Schneider}, \citenamefont {Turchi},\ and\ \citenamefont
  {Xiong}}]{Palumbo2014}%
  \BibitemOpen
  \bibfield  {author} {\bibinfo {author} {\bibfnamefont {M.}~\bibnamefont
  {Palumbo}}, \bibinfo {author} {\bibfnamefont {B.}~\bibnamefont {Burton}},
  \bibinfo {author} {\bibfnamefont {A.}~\bibnamefont {Costa~e silva}}, \bibinfo
  {author} {\bibfnamefont {B.}~\bibnamefont {Fultz}}, \bibinfo {author}
  {\bibfnamefont {B.}~\bibnamefont {Grabowski}}, \bibinfo {author}
  {\bibfnamefont {G.}~\bibnamefont {Grimvall}}, \bibinfo {author}
  {\bibfnamefont {B.}~\bibnamefont {Hallstedt}}, \bibinfo {author}
  {\bibfnamefont {O.}~\bibnamefont {Hellman}}, \bibinfo {author} {\bibfnamefont
  {B.}~\bibnamefont {Lindahl}}, \bibinfo {author} {\bibfnamefont
  {A.}~\bibnamefont {Schneider}}, \bibinfo {author} {\bibfnamefont {P.~E.}\
  \bibnamefont {Turchi}},\ and\ \bibinfo {author} {\bibfnamefont
  {W.}~\bibnamefont {Xiong}},\ }\bibfield  {title} {\bibinfo {title}
  {{Thermodynamic modelling of crystalline unary phases}},\ }\href
  {https://doi.org/10.1002/pssb.201350133} {\bibfield  {journal} {\bibinfo
  {journal} {Physica Status Solidi (B) Basic Research}\ }\textbf {\bibinfo
  {volume} {251}},\ \bibinfo {pages} {14} (\bibinfo {year} {2014})}\BibitemShut
  {NoStop}%
\bibitem [{\citenamefont {Moustafa}\ \emph {et~al.}(2017)\citenamefont
  {Moustafa}, \citenamefont {Schultz}, \citenamefont {Zurek},\ and\
  \citenamefont {Kofke}}]{Moustafa2017}%
  \BibitemOpen
  \bibfield  {author} {\bibinfo {author} {\bibfnamefont {S.~G.}\ \bibnamefont
  {Moustafa}}, \bibinfo {author} {\bibfnamefont {A.~J.}\ \bibnamefont
  {Schultz}}, \bibinfo {author} {\bibfnamefont {E.}~\bibnamefont {Zurek}},\
  and\ \bibinfo {author} {\bibfnamefont {D.~A.}\ \bibnamefont {Kofke}},\
  }\bibfield  {title} {\bibinfo {title} {{Accurate and precise ab initio
  anharmonic free-energy calculations for metallic crystals: Application to hcp
  Fe at high temperature and pressure}},\ }\href
  {https://doi.org/10.1103/PhysRevB.96.014117} {\bibfield  {journal} {\bibinfo
  {journal} {Physical Review B}\ }\textbf {\bibinfo {volume} {96}},\ \bibinfo
  {pages} {014117} (\bibinfo {year} {2017})}\BibitemShut {NoStop}%
\bibitem [{\citenamefont {Wagner}\ \emph {et~al.}(1998)\citenamefont {Wagner},
  \citenamefont {Laloyaux},\ and\ \citenamefont {Scheffler}}]{Wagner1998}%
  \BibitemOpen
  \bibfield  {author} {\bibinfo {author} {\bibfnamefont {F.}~\bibnamefont
  {Wagner}}, \bibinfo {author} {\bibfnamefont {T.}~\bibnamefont {Laloyaux}},\
  and\ \bibinfo {author} {\bibfnamefont {M.}~\bibnamefont {Scheffler}},\
  }\bibfield  {title} {\bibinfo {title} {{Errors in Hellmann-Feynman forces due
  to occupation-number broadening and how they can be corrected}},\ }\href
  {https://doi.org/10.1103/PhysRevB.57.2102} {\bibfield  {journal} {\bibinfo
  {journal} {Physical Review B - Condensed Matter and Materials Physics}\
  }\textbf {\bibinfo {volume} {57}},\ \bibinfo {pages} {2102} (\bibinfo {year}
  {1998})}\BibitemShut {NoStop}%
\bibitem [{\citenamefont {Kratzer}\ and\ \citenamefont
  {Neugebauer}(2019)}]{Kratzer2019}%
  \BibitemOpen
  \bibfield  {author} {\bibinfo {author} {\bibfnamefont {P.}~\bibnamefont
  {Kratzer}}\ and\ \bibinfo {author} {\bibfnamefont {J.}~\bibnamefont
  {Neugebauer}},\ }\bibfield  {title} {\bibinfo {title} {{The basics of
  electronic structure theory for periodic systems}},\ }\href
  {https://doi.org/10.3389/fchem.2019.00106} {\bibfield  {journal} {\bibinfo
  {journal} {Frontiers in Chemistry}\ }\textbf {\bibinfo {volume} {7}},\
  \bibinfo {pages} {1} (\bibinfo {year} {2019})}\BibitemShut {NoStop}%
\bibitem [{\citenamefont {Dove}(2011)}]{Dove2011}%
  \BibitemOpen
  \bibfield  {author} {\bibinfo {author} {\bibfnamefont {M.}~\bibnamefont
  {Dove}},\ }\bibfield  {title} {\bibinfo {title} {{Introduction to the theory
  of lattice dynamics}},\ }\href {https://doi.org/10.1051/sfn/201112007}
  {\bibfield  {journal} {\bibinfo  {journal} {{\'{E}}cole th{\'{e}}matique de
  la Soci{\'{e}}t{\'{e}} Fran{\c{c}}aise de la Neutronique}\ }\textbf {\bibinfo
  {volume} {12}},\ \bibinfo {pages} {123} (\bibinfo {year} {2011})}\BibitemShut
  {NoStop}%
\bibitem [{\citenamefont {Togo}\ and\ \citenamefont {Tanaka}(2015)}]{Togo2015}%
  \BibitemOpen
  \bibfield  {author} {\bibinfo {author} {\bibfnamefont {A.}~\bibnamefont
  {Togo}}\ and\ \bibinfo {author} {\bibfnamefont {I.}~\bibnamefont {Tanaka}},\
  }\bibfield  {title} {\bibinfo {title} {{First principles phonon calculations
  in materials science}},\ }\href
  {https://doi.org/10.1016/j.scriptamat.2015.07.021} {\bibfield  {journal}
  {\bibinfo  {journal} {Scripta Materialia}\ }\textbf {\bibinfo {volume}
  {108}},\ \bibinfo {pages} {1} (\bibinfo {year} {2015})}\BibitemShut {NoStop}%
\bibitem [{\citenamefont {Frenkel}\ and\ \citenamefont {Smit}(2002)}]{Frenkel}%
  \BibitemOpen
  \bibfield  {author} {\bibinfo {author} {\bibfnamefont {D.}~\bibnamefont
  {Frenkel}}\ and\ \bibinfo {author} {\bibfnamefont {B.}~\bibnamefont {Smit}},\
  }\href {https://doi.org/10.1016/B978-0-12-267351-1.X5000-7} {\emph {\bibinfo
  {title} {{Understanding Molecular Simulation}}}}\ (\bibinfo  {publisher}
  {Academic Press},\ \bibinfo {address} {San Diego},\ \bibinfo {year}
  {2002})\BibitemShut {NoStop}%
\bibitem [{\citenamefont {Rickman}\ and\ \citenamefont
  {LeSar}(2002)}]{Rickman2002}%
  \BibitemOpen
  \bibfield  {author} {\bibinfo {author} {\bibfnamefont {J.~M.}\ \bibnamefont
  {Rickman}}\ and\ \bibinfo {author} {\bibfnamefont {R.}~\bibnamefont
  {LeSar}},\ }\bibfield  {title} {\bibinfo {title} {{Free-energy calculations
  in materials research}},\ }\href
  {https://doi.org/10.1146/annurev.matsci.32.111901.153708} {\bibfield
  {journal} {\bibinfo  {journal} {Annual Review of Materials Science}\ }\textbf
  {\bibinfo {volume} {32}},\ \bibinfo {pages} {195} (\bibinfo {year}
  {2002})}\BibitemShut {NoStop}%
\bibitem [{\citenamefont {Chipot}\ and\ \citenamefont
  {Pohorille}(2007)}]{Chipot2007}%
  \BibitemOpen
  \bibfield  {author} {\bibinfo {author} {\bibfnamefont {C.}~\bibnamefont
  {Chipot}}\ and\ \bibinfo {author} {\bibfnamefont {A.}~\bibnamefont
  {Pohorille}},\ }\bibfield  {title} {\bibinfo {title} {{Calculating Free
  Energy Differences Using Perturbation Theory}},\ }in\ \href
  {https://doi.org/10.1007/978-3-540-38448-9_2} {\emph {\bibinfo {booktitle}
  {Free Energy Calculations}}},\ Vol.~\bibinfo {volume} {86},\ \bibinfo
  {editor} {edited by\ \bibinfo {editor} {\bibfnamefont {A.}~\bibnamefont
  {Pohorille}}\ and\ \bibinfo {editor} {\bibfnamefont {C.}~\bibnamefont
  {Chipot}}}\ (\bibinfo  {publisher} {Springer},\ \bibinfo {address} {Berlin,
  Heidelberg},\ \bibinfo {year} {2007})\ Chap.~\bibinfo {chapter} {2}, pp.\
  \bibinfo {pages} {33--75}\BibitemShut {NoStop}%
\bibitem [{\citenamefont {Vo{\v{c}}adlo}\ and\ \citenamefont
  {Alf{\`{e}}}(2002)}]{Vocadlo2002}%
  \BibitemOpen
  \bibfield  {author} {\bibinfo {author} {\bibfnamefont {L.}~\bibnamefont
  {Vo{\v{c}}adlo}}\ and\ \bibinfo {author} {\bibfnamefont {D.}~\bibnamefont
  {Alf{\`{e}}}},\ }\bibfield  {title} {\bibinfo {title} {{Ab initio melting
  curve of the fcc phase of aluminum}},\ }\href
  {https://doi.org/10.1103/PhysRevB.65.214105} {\bibfield  {journal} {\bibinfo
  {journal} {Physical Review B - Condensed Matter and Materials Physics}\
  }\textbf {\bibinfo {volume} {65}},\ \bibinfo {pages} {1} (\bibinfo {year}
  {2002})}\BibitemShut {NoStop}%
\bibitem [{\citenamefont {Grabowski}\ \emph {et~al.}(2009)\citenamefont
  {Grabowski}, \citenamefont {Ismer}, \citenamefont {Hickel},\ and\
  \citenamefont {Neugebauer}}]{Grabowski2009}%
  \BibitemOpen
  \bibfield  {author} {\bibinfo {author} {\bibfnamefont {B.}~\bibnamefont
  {Grabowski}}, \bibinfo {author} {\bibfnamefont {L.}~\bibnamefont {Ismer}},
  \bibinfo {author} {\bibfnamefont {T.}~\bibnamefont {Hickel}},\ and\ \bibinfo
  {author} {\bibfnamefont {J.}~\bibnamefont {Neugebauer}},\ }\bibfield  {title}
  {\bibinfo {title} {{Ab initio up to the melting point: Anharmonicity and
  vacancies in aluminum}},\ }\href {https://doi.org/10.1103/PhysRevB.79.134106}
  {\bibfield  {journal} {\bibinfo  {journal} {Physical Review B - Condensed
  Matter and Materials Physics}\ }\textbf {\bibinfo {volume} {79}},\ \bibinfo
  {pages} {134106} (\bibinfo {year} {2009})}\BibitemShut {NoStop}%
\bibitem [{\citenamefont {Duff}\ \emph {et~al.}(2015)\citenamefont {Duff},
  \citenamefont {Davey}, \citenamefont {Korbmacher}, \citenamefont {Glensk},
  \citenamefont {Grabowski}, \citenamefont {Neugebauer},\ and\ \citenamefont
  {Finnis}}]{Duff2015}%
  \BibitemOpen
  \bibfield  {author} {\bibinfo {author} {\bibfnamefont {A.~I.}\ \bibnamefont
  {Duff}}, \bibinfo {author} {\bibfnamefont {T.}~\bibnamefont {Davey}},
  \bibinfo {author} {\bibfnamefont {D.}~\bibnamefont {Korbmacher}}, \bibinfo
  {author} {\bibfnamefont {A.}~\bibnamefont {Glensk}}, \bibinfo {author}
  {\bibfnamefont {B.}~\bibnamefont {Grabowski}}, \bibinfo {author}
  {\bibfnamefont {J.}~\bibnamefont {Neugebauer}},\ and\ \bibinfo {author}
  {\bibfnamefont {M.~W.}\ \bibnamefont {Finnis}},\ }\bibfield  {title}
  {\bibinfo {title} {{Improved method of calculating ab initio high-temperature
  thermodynamic properties with application to ZrC}},\ }\href
  {https://doi.org/10.1103/PhysRevB.91.214311} {\bibfield  {journal} {\bibinfo
  {journal} {Physical Review B - Condensed Matter and Materials Physics}\
  }\textbf {\bibinfo {volume} {91}},\ \bibinfo {pages} {214311} (\bibinfo
  {year} {2015})}\BibitemShut {NoStop}%
\bibitem [{\citenamefont {Sun}\ \emph {et~al.}(2018)\citenamefont {Sun},
  \citenamefont {Brodholt}, \citenamefont {Li},\ and\ \citenamefont
  {Vo{\v{c}}adlo}}]{Sun2018}%
  \BibitemOpen
  \bibfield  {author} {\bibinfo {author} {\bibfnamefont {T.}~\bibnamefont
  {Sun}}, \bibinfo {author} {\bibfnamefont {J.~P.}\ \bibnamefont {Brodholt}},
  \bibinfo {author} {\bibfnamefont {Y.}~\bibnamefont {Li}},\ and\ \bibinfo
  {author} {\bibfnamefont {L.}~\bibnamefont {Vo{\v{c}}adlo}},\ }\bibfield
  {title} {\bibinfo {title} {{Melting properties from ab initio free energy
  calculations: Iron at the Earth's inner-core boundary}},\ }\href
  {https://doi.org/10.1103/PhysRevB.98.224301} {\bibfield  {journal} {\bibinfo
  {journal} {Physical Review B}\ }\textbf {\bibinfo {volume} {98}},\ \bibinfo
  {pages} {1} (\bibinfo {year} {2018})}\BibitemShut {NoStop}%
\bibitem [{\citenamefont {Rang}\ and\ \citenamefont {Kresse}(2019)}]{Rang2019}%
  \BibitemOpen
  \bibfield  {author} {\bibinfo {author} {\bibfnamefont {M.}~\bibnamefont
  {Rang}}\ and\ \bibinfo {author} {\bibfnamefont {G.}~\bibnamefont {Kresse}},\
  }\bibfield  {title} {\bibinfo {title} {{First-principles study of the melting
  temperature of MgO}},\ }\href {https://doi.org/10.1103/PhysRevB.99.184103}
  {\bibfield  {journal} {\bibinfo  {journal} {Physical Review B}\ }\textbf
  {\bibinfo {volume} {99}},\ \bibinfo {pages} {184103} (\bibinfo {year}
  {2019})}\BibitemShut {NoStop}%
\bibitem [{\citenamefont {Frenkel}\ and\ \citenamefont
  {Ladd}(1984)}]{Frenkel1984}%
  \BibitemOpen
  \bibfield  {author} {\bibinfo {author} {\bibfnamefont {D.}~\bibnamefont
  {Frenkel}}\ and\ \bibinfo {author} {\bibfnamefont {A.~J.}\ \bibnamefont
  {Ladd}},\ }\bibfield  {title} {\bibinfo {title} {{New Monte Carlo method to
  compute the free energy of arbitrary solids. Application to the fcc and hcp
  phases of hard spheres}},\ }\href {https://doi.org/10.1063/1.448024}
  {\bibfield  {journal} {\bibinfo  {journal} {The Journal of Chemical Physics}\
  }\textbf {\bibinfo {volume} {81}},\ \bibinfo {pages} {3188} (\bibinfo {year}
  {1984})}\BibitemShut {NoStop}%
\bibitem [{\citenamefont {Kirkwood}(1935)}]{Kirkwood1935}%
  \BibitemOpen
  \bibfield  {author} {\bibinfo {author} {\bibfnamefont {J.~G.}\ \bibnamefont
  {Kirkwood}},\ }\bibfield  {title} {\bibinfo {title} {{Statistical Mechanics
  of Fluid Mixtures}},\ }\href {https://doi.org/10.1063/1.1749657} {\bibfield
  {journal} {\bibinfo  {journal} {The Journal of Chemical Physics}\ }\textbf
  {\bibinfo {volume} {3}},\ \bibinfo {pages} {300} (\bibinfo {year}
  {1935})}\BibitemShut {NoStop}%
\bibitem [{\citenamefont {Wood}\ \emph {et~al.}(1991)\citenamefont {Wood},
  \citenamefont {Muhlbauer},\ and\ \citenamefont {Thompson}}]{Wood1991}%
  \BibitemOpen
  \bibfield  {author} {\bibinfo {author} {\bibfnamefont {R.~H.}\ \bibnamefont
  {Wood}}, \bibinfo {author} {\bibfnamefont {W.~C.~F.}\ \bibnamefont
  {Muhlbauer}},\ and\ \bibinfo {author} {\bibfnamefont {P.~T.}\ \bibnamefont
  {Thompson}},\ }\bibfield  {title} {\bibinfo {title} {{Systematic errors in
  free energy perturbation calculations due to a finite sample of configuration
  space: sample-size hysteresis}},\ }\href
  {https://doi.org/10.1021/j100170a054} {\bibfield  {journal} {\bibinfo
  {journal} {The Journal of Physical Chemistry}\ }\textbf {\bibinfo {volume}
  {95}},\ \bibinfo {pages} {6670} (\bibinfo {year} {1991})}\BibitemShut
  {NoStop}%
\bibitem [{\citenamefont {Zuckerman}\ and\ \citenamefont
  {Woolf}(2002)}]{Zuckerman2002}%
  \BibitemOpen
  \bibfield  {author} {\bibinfo {author} {\bibfnamefont {D.~M.}\ \bibnamefont
  {Zuckerman}}\ and\ \bibinfo {author} {\bibfnamefont {T.~B.}\ \bibnamefont
  {Woolf}},\ }\bibfield  {title} {\bibinfo {title} {{Theory of a Systematic
  Computational Error in Free Energy Differences}},\ }\href
  {https://doi.org/10.1103/PhysRevLett.89.180602} {\bibfield  {journal}
  {\bibinfo  {journal} {Physical Review Letters}\ }\textbf {\bibinfo {volume}
  {89}},\ \bibinfo {pages} {180602} (\bibinfo {year} {2002})}\BibitemShut
  {NoStop}%
\bibitem [{\citenamefont {Gore}\ \emph {et~al.}(2003)\citenamefont {Gore},
  \citenamefont {Ritort},\ and\ \citenamefont {Bustamante}}]{Gore2003}%
  \BibitemOpen
  \bibfield  {author} {\bibinfo {author} {\bibfnamefont {J.}~\bibnamefont
  {Gore}}, \bibinfo {author} {\bibfnamefont {F.}~\bibnamefont {Ritort}},\ and\
  \bibinfo {author} {\bibfnamefont {C.}~\bibnamefont {Bustamante}},\ }\bibfield
   {title} {\bibinfo {title} {{Bias and error in estimates of equilibrium
  free-energy differences from nonequilibrium measurements}},\ }\href
  {https://doi.org/10.1073/pnas.1635159100} {\bibfield  {journal} {\bibinfo
  {journal} {Proceedings of the National Academy of Sciences}\ }\textbf
  {\bibinfo {volume} {100}},\ \bibinfo {pages} {12564} (\bibinfo {year}
  {2003})}\BibitemShut {NoStop}%
\bibitem [{\citenamefont {Zuckerman}\ and\ \citenamefont
  {Woolf}(2004)}]{Zuckerman2004}%
  \BibitemOpen
  \bibfield  {author} {\bibinfo {author} {\bibfnamefont {D.~M.}\ \bibnamefont
  {Zuckerman}}\ and\ \bibinfo {author} {\bibfnamefont {T.~B.}\ \bibnamefont
  {Woolf}},\ }\bibfield  {title} {\bibinfo {title} {{Systematic Finite-Sampling
  Inaccuracy in Free Energy Differences and Other Nonlinear Quantities}},\
  }\href {https://doi.org/10.1023/B:JOSS.0000013961.84860.5b} {\bibfield
  {journal} {\bibinfo  {journal} {Journal of Statistical Physics}\ }\textbf
  {\bibinfo {volume} {114}},\ \bibinfo {pages} {1303} (\bibinfo {year}
  {2004})}\BibitemShut {NoStop}%
\bibitem [{\citenamefont {Wu}\ and\ \citenamefont {Kofke}(2004)}]{Wu2004}%
  \BibitemOpen
  \bibfield  {author} {\bibinfo {author} {\bibfnamefont {D.}~\bibnamefont
  {Wu}}\ and\ \bibinfo {author} {\bibfnamefont {D.~A.}\ \bibnamefont {Kofke}},\
  }\bibfield  {title} {\bibinfo {title} {{Model for small-sample bias of
  free-energy calculations applied to Gaussian-distributed nonequilibrium work
  measurements}},\ }\href {https://doi.org/10.1063/1.1806413} {\bibfield
  {journal} {\bibinfo  {journal} {The Journal of Chemical Physics}\ }\textbf
  {\bibinfo {volume} {121}},\ \bibinfo {pages} {8742} (\bibinfo {year}
  {2004})}\BibitemShut {NoStop}%
\bibitem [{\citenamefont {Wu}\ and\ \citenamefont {Kofke}(2005)}]{Wu2005}%
  \BibitemOpen
  \bibfield  {author} {\bibinfo {author} {\bibfnamefont {D.}~\bibnamefont
  {Wu}}\ and\ \bibinfo {author} {\bibfnamefont {D.~A.}\ \bibnamefont {Kofke}},\
  }\bibfield  {title} {\bibinfo {title} {{Phase-space overlap measures. I.
  Fail-safe bias detection in free energies calculated by molecular
  simulation}},\ }\bibfield  {journal} {\bibinfo  {journal} {Journal of
  Chemical Physics}\ }\textbf {\bibinfo {volume} {123}},\ \href
  {https://doi.org/10.1063/1.1992483} {10.1063/1.1992483} (\bibinfo {year}
  {2005})\BibitemShut {NoStop}%
\bibitem [{\citenamefont {Jarzynski}(2006)}]{Jarzynski2006}%
  \BibitemOpen
  \bibfield  {author} {\bibinfo {author} {\bibfnamefont {C.}~\bibnamefont
  {Jarzynski}},\ }\bibfield  {title} {\bibinfo {title} {{Rare events and the
  convergence of exponentially averaged work values}},\ }\href
  {https://doi.org/10.1103/PhysRevE.73.046105} {\bibfield  {journal} {\bibinfo
  {journal} {Physical Review E}\ }\textbf {\bibinfo {volume} {73}},\ \bibinfo
  {pages} {046105} (\bibinfo {year} {2006})}\BibitemShut {NoStop}%
\bibitem [{\citenamefont {Pohorille}\ \emph {et~al.}(2010)\citenamefont
  {Pohorille}, \citenamefont {Jarzynski},\ and\ \citenamefont
  {Chipot}}]{Pohorille2010}%
  \BibitemOpen
  \bibfield  {author} {\bibinfo {author} {\bibfnamefont {A.}~\bibnamefont
  {Pohorille}}, \bibinfo {author} {\bibfnamefont {C.}~\bibnamefont
  {Jarzynski}},\ and\ \bibinfo {author} {\bibfnamefont {C.}~\bibnamefont
  {Chipot}},\ }\bibfield  {title} {\bibinfo {title} {{Good practices in
  free-energy calculations}},\ }\href {https://doi.org/10.1021/jp102971x}
  {\bibfield  {journal} {\bibinfo  {journal} {Journal of Physical Chemistry B}\
  }\textbf {\bibinfo {volume} {114}},\ \bibinfo {pages} {10235} (\bibinfo
  {year} {2010})}\BibitemShut {NoStop}%
\bibitem [{\citenamefont {Boresch}\ and\ \citenamefont
  {Woodcock}(2017)}]{Boresch2017}%
  \BibitemOpen
  \bibfield  {author} {\bibinfo {author} {\bibfnamefont {S.}~\bibnamefont
  {Boresch}}\ and\ \bibinfo {author} {\bibfnamefont {H.~L.}\ \bibnamefont
  {Woodcock}},\ }\bibfield  {title} {\bibinfo {title} {{Convergence of
  single-step free energy perturbation}},\ }\href
  {https://doi.org/10.1080/00268976.2016.1269960} {\bibfield  {journal}
  {\bibinfo  {journal} {Molecular Physics}\ }\textbf {\bibinfo {volume}
  {115}},\ \bibinfo {pages} {1200} (\bibinfo {year} {2017})}\BibitemShut
  {NoStop}%
\bibitem [{\citenamefont {Wood}\ and\ \citenamefont
  {Thompson}(2018)}]{Wood2018}%
  \BibitemOpen
  \bibfield  {author} {\bibinfo {author} {\bibfnamefont {M.~A.}\ \bibnamefont
  {Wood}}\ and\ \bibinfo {author} {\bibfnamefont {A.~P.}\ \bibnamefont
  {Thompson}},\ }\bibfield  {title} {\bibinfo {title} {{Extending the accuracy
  of the SNAP interatomic potential form}},\ }\href
  {https://doi.org/10.1063/1.5017641} {\bibfield  {journal} {\bibinfo
  {journal} {The Journal of Chemical Physics}\ }\textbf {\bibinfo {volume}
  {148}},\ \bibinfo {pages} {241721} (\bibinfo {year} {2018})}\BibitemShut
  {NoStop}%
\bibitem [{\citenamefont {Shyu}\ and\ \citenamefont
  {Ytreberg}(2009)}]{Shyu2009}%
  \BibitemOpen
  \bibfield  {author} {\bibinfo {author} {\bibfnamefont {C.}~\bibnamefont
  {Shyu}}\ and\ \bibinfo {author} {\bibfnamefont {F.~M.}\ \bibnamefont
  {Ytreberg}},\ }\bibfield  {title} {\bibinfo {title} {{Reducing the bias and
  uncertainty of free energy estimates by using regression to fit thermodynamic
  integration data}},\ }\href {https://doi.org/10.1002/jcc.21231} {\bibfield
  {journal} {\bibinfo  {journal} {Journal of Computational Chemistry}\ ,\
  \bibinfo {pages} {NA}} (\bibinfo {year} {2009})}\BibitemShut {NoStop}%
\bibitem [{\citenamefont {Jorge}\ \emph {et~al.}(2010)\citenamefont {Jorge},
  \citenamefont {Garrido}, \citenamefont {Queimada}, \citenamefont {Economou},\
  and\ \citenamefont {MacEdo}}]{Jorge2010}%
  \BibitemOpen
  \bibfield  {author} {\bibinfo {author} {\bibfnamefont {M.}~\bibnamefont
  {Jorge}}, \bibinfo {author} {\bibfnamefont {N.~M.}\ \bibnamefont {Garrido}},
  \bibinfo {author} {\bibfnamefont {A.~J.}\ \bibnamefont {Queimada}}, \bibinfo
  {author} {\bibfnamefont {I.~G.}\ \bibnamefont {Economou}},\ and\ \bibinfo
  {author} {\bibfnamefont {E.~A.}\ \bibnamefont {MacEdo}},\ }\bibfield  {title}
  {\bibinfo {title} {{Effect of the integration method on the accuracy and
  computational efficiency of free energy calculations using thermodynamic
  integration}},\ }\href {https://doi.org/10.1021/ct900661c} {\bibfield
  {journal} {\bibinfo  {journal} {Journal of Chemical Theory and Computation}\
  }\textbf {\bibinfo {volume} {6}},\ \bibinfo {pages} {1018} (\bibinfo {year}
  {2010})}\BibitemShut {NoStop}%
\bibitem [{\citenamefont {Ryde}(2017)}]{Ryde2017}%
  \BibitemOpen
  \bibfield  {author} {\bibinfo {author} {\bibfnamefont {U.}~\bibnamefont
  {Ryde}},\ }\bibfield  {title} {\bibinfo {title} {{How Many Conformations Need
  to Be Sampled to Obtain Converged QM/MM Energies? the Curse of Exponential
  Averaging}},\ }\href {https://doi.org/10.1021/acs.jctc.7b00826} {\bibfield
  {journal} {\bibinfo  {journal} {Journal of Chemical Theory and Computation}\
  }\textbf {\bibinfo {volume} {13}},\ \bibinfo {pages} {5745} (\bibinfo {year}
  {2017})}\BibitemShut {NoStop}%
\bibitem [{\citenamefont {Hellman}(2012)}]{Hellman2012}%
  \BibitemOpen
  \bibfield  {author} {\bibinfo {author} {\bibfnamefont {O.}~\bibnamefont
  {Hellman}},\ }\emph {\bibinfo {title} {{Thermal properties of materials from
  first principles}}},\ \href@noop {} {Ph.D. thesis},\ \bibinfo  {school}
  {Link{\"{o}}ping University} (\bibinfo {year} {2012})\BibitemShut {NoStop}%
\bibitem [{\citenamefont {Efron}(1979)}]{Efron1979}%
  \BibitemOpen
  \bibfield  {author} {\bibinfo {author} {\bibfnamefont {B.}~\bibnamefont
  {Efron}},\ }\bibfield  {title} {\bibinfo {title} {{Bootstrap Methods: Another
  Look at the Jackknife}},\ }\href {https://doi.org/10.1214/aos/1176344552}
  {\bibfield  {journal} {\bibinfo  {journal} {The Annals of Statistics}\
  }\textbf {\bibinfo {volume} {7}},\ \bibinfo {pages} {1100} (\bibinfo {year}
  {1979})}\BibitemShut {NoStop}%
\bibitem [{\citenamefont {Kresse}\ and\ \citenamefont
  {Hafner}(1993)}]{Kresse1993}%
  \BibitemOpen
  \bibfield  {author} {\bibinfo {author} {\bibfnamefont {G.}~\bibnamefont
  {Kresse}}\ and\ \bibinfo {author} {\bibfnamefont {J.}~\bibnamefont
  {Hafner}},\ }\bibfield  {title} {\bibinfo {title} {{Ab initio molecular
  dynamics for liquid metals}},\ }\href@noop {} {\bibfield  {journal} {\bibinfo
   {journal} {Physical Review B}\ }\textbf {\bibinfo {volume} {47}} (\bibinfo
  {year} {1993})}\BibitemShut {NoStop}%
\bibitem [{\citenamefont {Kresse}\ and\ \citenamefont
  {Hafner}(1994)}]{Kresse1994a}%
  \BibitemOpen
  \bibfield  {author} {\bibinfo {author} {\bibfnamefont {G.}~\bibnamefont
  {Kresse}}\ and\ \bibinfo {author} {\bibfnamefont {J.}~\bibnamefont
  {Hafner}},\ }\bibfield  {title} {\bibinfo {title} {{Ab initio
  molecular-dynamics simulation of the liquid-metal-amorphous-semiconductor
  transition in germanium}},\ }\href@noop {} {\bibfield  {journal} {\bibinfo
  {journal} {Phys. Rev. B}\ }\textbf {\bibinfo {volume} {49}},\ \bibinfo
  {pages} {14251} (\bibinfo {year} {1994})}\BibitemShut {NoStop}%
\bibitem [{\citenamefont {Kresse}\ and\ \citenamefont
  {Furthmuller}(1996)}]{Kresse1996}%
  \BibitemOpen
  \bibfield  {author} {\bibinfo {author} {\bibfnamefont {G.}~\bibnamefont
  {Kresse}}\ and\ \bibinfo {author} {\bibfnamefont {J.}~\bibnamefont
  {Furthmuller}},\ }\bibfield  {title} {\bibinfo {title} {{Efficiency of
  ab-initio total energy calculations for metals and semiconductors using a
  plane-wave basis set}},\ }\href
  {https://doi.org/10.1016/0927-0256(96)00008-0} {\bibfield  {journal}
  {\bibinfo  {journal} {Comp. Mater. Sci.}\ }\textbf {\bibinfo {volume} {6}},\
  \bibinfo {pages} {15} (\bibinfo {year} {1996})}\BibitemShut {NoStop}%
\bibitem [{\citenamefont {Kresse}\ and\ \citenamefont
  {Furthm\"{u}ller}(1996)}]{Kresse1996a}%
  \BibitemOpen
  \bibfield  {author} {\bibinfo {author} {\bibfnamefont {G.}~\bibnamefont
  {Kresse}}\ and\ \bibinfo {author} {\bibfnamefont {J.}~\bibnamefont
  {Furthm\"{u}ller}},\ }\bibfield  {title} {\bibinfo {title} {{Efficient
  iterative schemes for ab initio total-energy calculations using a plane-wave
  basis set.}},\ }\href@noop {} {\bibfield  {journal} {\bibinfo  {journal}
  {Physical review. B, Condensed matter}\ }\textbf {\bibinfo {volume} {54}},\
  \bibinfo {pages} {11169} (\bibinfo {year} {1996})}\BibitemShut {NoStop}%
\bibitem [{\citenamefont {Kresse}\ and\ \citenamefont
  {Joubert}(1999)}]{Kresse1999}%
  \BibitemOpen
  \bibfield  {author} {\bibinfo {author} {\bibfnamefont {G.}~\bibnamefont
  {Kresse}}\ and\ \bibinfo {author} {\bibfnamefont {D.}~\bibnamefont
  {Joubert}},\ }\bibfield  {title} {\bibinfo {title} {{From ultrasoft
  pseudopotentials to the projector augmented-wave method}},\ }\href
  {https://doi.org/10.1103/PhysRevB.59.1758} {\bibfield  {journal} {\bibinfo
  {journal} {Phys. Rev. B}\ }\textbf {\bibinfo {volume} {59}},\ \bibinfo
  {pages} {1758} (\bibinfo {year} {1999})}\BibitemShut {NoStop}%
\bibitem [{fit()}]{fitsnap}%
  \BibitemOpen
  \href {https://github.com/FitSNAP/FitSNAP} {\bibinfo {title} {{FitSNAP.
  https://github.com/FitSNAP/FitSNAP}}}\BibitemShut {NoStop}%
\bibitem [{\citenamefont {Hellman}\ \emph {et~al.}(2011)\citenamefont
  {Hellman}, \citenamefont {Abrikosov},\ and\ \citenamefont
  {Simak}}]{Hellman2011}%
  \BibitemOpen
  \bibfield  {author} {\bibinfo {author} {\bibfnamefont {O.}~\bibnamefont
  {Hellman}}, \bibinfo {author} {\bibfnamefont {I.~A.}\ \bibnamefont
  {Abrikosov}},\ and\ \bibinfo {author} {\bibfnamefont {S.~I.}\ \bibnamefont
  {Simak}},\ }\bibfield  {title} {\bibinfo {title} {{Lattice dynamics of
  anharmonic solids from first principles}},\ }\href
  {https://doi.org/10.1103/PhysRevB.84.180301} {\bibfield  {journal} {\bibinfo
  {journal} {Physical Review B}\ }\textbf {\bibinfo {volume} {84}},\ \bibinfo
  {pages} {180301(R)} (\bibinfo {year} {2011})}\BibitemShut {NoStop}%
\bibitem [{\citenamefont {Thompson}\ \emph {et~al.}(2015)\citenamefont
  {Thompson}, \citenamefont {Swiler}, \citenamefont {Trott}, \citenamefont
  {Foiles},\ and\ \citenamefont {Tucker}}]{Thompson2014}%
  \BibitemOpen
  \bibfield  {author} {\bibinfo {author} {\bibfnamefont {A.}~\bibnamefont
  {Thompson}}, \bibinfo {author} {\bibfnamefont {L.}~\bibnamefont {Swiler}},
  \bibinfo {author} {\bibfnamefont {C.}~\bibnamefont {Trott}}, \bibinfo
  {author} {\bibfnamefont {S.}~\bibnamefont {Foiles}},\ and\ \bibinfo {author}
  {\bibfnamefont {G.}~\bibnamefont {Tucker}},\ }\bibfield  {title} {\bibinfo
  {title} {{Spectral neighbor analysis method for automated generation of
  quantum-accurate interatomic potentials}},\ }\href
  {https://doi.org/10.1016/j.jcp.2014.12.018} {\bibfield  {journal} {\bibinfo
  {journal} {Journal of Computational Physics}\ }\textbf {\bibinfo {volume}
  {285}},\ \bibinfo {pages} {316} (\bibinfo {year} {2015})}\BibitemShut
  {NoStop}%
\bibitem [{\citenamefont {Plimpton}(1995)}]{Plimpton1997}%
  \BibitemOpen
  \bibfield  {author} {\bibinfo {author} {\bibfnamefont {S.}~\bibnamefont
  {Plimpton}},\ }\bibfield  {title} {\bibinfo {title} {{Fast Parallel
  Algorithms for Short-Range Molecular Dynamics}},\ }\href
  {https://doi.org/10.1006/jcph.1995.1039} {\bibfield  {journal} {\bibinfo
  {journal} {Journal of Computational Physics}\ }\textbf {\bibinfo {volume}
  {117}},\ \bibinfo {pages} {1} (\bibinfo {year} {1995})}\BibitemShut {NoStop}%
\bibitem [{\citenamefont {Ceriotti}\ \emph {et~al.}(2009)\citenamefont
  {Ceriotti}, \citenamefont {Bussi},\ and\ \citenamefont
  {Parrinello}}]{Ceriotti2009}%
  \BibitemOpen
  \bibfield  {author} {\bibinfo {author} {\bibfnamefont {M.}~\bibnamefont
  {Ceriotti}}, \bibinfo {author} {\bibfnamefont {G.}~\bibnamefont {Bussi}},\
  and\ \bibinfo {author} {\bibfnamefont {M.}~\bibnamefont {Parrinello}},\
  }\bibfield  {title} {\bibinfo {title} {{Langevin equation with colored noise
  for constant-temperature molecular dynamics simulations}},\ }\href
  {https://doi.org/10.1103/PhysRevLett.102.020601} {\bibfield  {journal}
  {\bibinfo  {journal} {Physical Review Letters}\ }\textbf {\bibinfo {volume}
  {102}},\ \bibinfo {pages} {1} (\bibinfo {year} {2009})}\BibitemShut {NoStop}%
\bibitem [{\citenamefont {Ceriotti}\ \emph {et~al.}(2010)\citenamefont
  {Ceriotti}, \citenamefont {Bussi},\ and\ \citenamefont
  {Parrinello}}]{Ceriotti2010a}%
  \BibitemOpen
  \bibfield  {author} {\bibinfo {author} {\bibfnamefont {M.}~\bibnamefont
  {Ceriotti}}, \bibinfo {author} {\bibfnamefont {G.}~\bibnamefont {Bussi}},\
  and\ \bibinfo {author} {\bibfnamefont {M.}~\bibnamefont {Parrinello}},\
  }\bibfield  {title} {\bibinfo {title} {{Colored-Noise Thermostats {\`{a}} la
  Carte}},\ }\href {https://doi.org/10.1021/ct900563s} {\bibfield  {journal}
  {\bibinfo  {journal} {Journal of Chemical Theory and Computation}\ }\textbf
  {\bibinfo {volume} {6}},\ \bibinfo {pages} {1170} (\bibinfo {year}
  {2010})}\BibitemShut {NoStop}%
\end{thebibliography}%

\end{document}